\documentclass[aps,prb,reprint,groupedaddress,amsmath,amssymb,floatfix]{revtex4-2}
\usepackage[T1]{fontenc}
\usepackage{graphicx}
\usepackage{tikz}
\usetikzlibrary{arrows.meta,calc,positioning,decorations.pathreplacing}
\usepackage{booktabs}
\usepackage{bm,braket}
\usepackage[hidelinks]{hyperref}
\graphicspath{{./}}

\let\OriginalBibitemNoStop\bibitemNoStop
\renewcommand{\bibitemNoStop}{\unskip\OriginalBibitemNoStop}

\newtheorem{thm}{Theorem}[section]
\newtheorem{prop}[thm]{Proposition}

\newtheorem{cor}[thm]{Corollary}

\newtheorem{dfn}[thm]{Definition}

\newcommand{\im}{{\rm im\,}}

\newcommand{\tr}{{\rm tr\,}}

\newcommand{\sgn}{{\rm sgn\,}}
\newcommand{\p}{\partial}

\newcommand{\s}{{\sigma}}

\newcommand{\C}{\mathbb{C}}

\newcommand{\R}{\mathbb{R}}
\newcommand{\Z}{\mathbb{Z}}

\newcommand{\bx}{{\bm{x}}}
\newcommand{\bfx}{{\mathbf{x}}}
\newcommand{\bk}{{\bm{k}}}
\newcommand{\br}{{\bm{r}}}

\newcommand{\ba}{{\bm{a}}}
\newcommand{\bt}{{\bm{t}}}
\newcommand{\bR}{{\bm{R}}}
\newcommand{\bG}{{\bm{G}}}
\newcommand{\bw}{{\bm{w}}}
\newcommand{\sCS}{{\rm sCS}}

\def\widebar{\accentset{{\cc@st_yle\underline{\mskip10mu}}}}
\def\wideubar{\underaccent{{\cc@st_yle\underline{\mskip10mu}}}}

\begin{document}
\title{Topology of Wannierizable Insulators}
\author{Ken Shiozaki}
\email{ken.shiozaki@yukawa.kyoto-u.ac.jp}
\affiliation{Center for Gravitational Physics and Quantum Information, Yukawa Institute for Theoretical Physics, Kyoto University, Kyoto 606-8502, Japan}
\date{\today}
\begin{abstract}
We study the freedom in choosing Wannier centers and the associated symmetry representations in noninteracting Wannierizable insulators.
We compare Wyckoff deformations of atomic configurations with symmetry-preserving adiabatic deformations of occupied states, both with and without adding common occupied atomic orbitals.
Explicit examples show that stable Wyckoff equivalence and adiabatic equivalence at fixed filling do not imply each other.
In spinless $Fm'm'm'$, an Euler class distinguishes stably Wyckoff-equivalent configurations.
In spinless $P\overline1'$, adiabatically equivalent configurations remain distinct under stable Wyckoff deformations.
We also classify when stable Wyckoff and stable adiabatic equivalence coincide.
They agree in two or fewer dimensions and in three-dimensional spinful systems.
For three-dimensional spinless systems, Chern insulator pumps give additional equivalences in 134 magnetic space groups.
These groups preserve spacetime orientation and have a point whose site-symmetry group contains $222$, $\overline1'$, or $\overline4'$.
\end{abstract}
\maketitle
%\tableofcontents

\section{Introduction}

Wannier states are a basic tool for describing crystalline electronic states in real space~\cite{MarzariMaximally2012}.
A fundamental question in the study of topological insulators is whether a symmetry-compatible, exponentially localized Wannier basis exists.
Here, compatibility with symmetry means that a symmetry operation maps the Wannier orbitals at a given center to linear combinations of orbitals at the symmetry-related center. 
A system admitting a symmetry-compatible, exponentially localized Wannier basis is called a Wannierizable insulator and can be adiabatically connected to an atomic limit. 
Throughout the paper, a Wannier basis is understood to be symmetry compatible and exponentially localized.

If a complete set of topological invariants distinguishing the isomorphism classes of occupied Bloch bundles were known, Wannierizability could in principle be determined by comparing these invariants with those of band representations induced from atomic limits.
However, complete sets of topological invariants are known only in limited settings.
For stable equivalence, invariants defined on the one-dimensional skeleton of the Brillouin zone (BZ) have been determined~\cite{OnoComplete2024}.

In this paper, we focus on Wannierizable insulators and ask how uniquely a Wannier basis is determined.

The spatial profiles of Wannier orbitals depend on the gauge of Bloch wave functions and may change under adiabatic deformations of the Hamiltonian.
We therefore focus on the Wyckoff positions of the Wannier centers and the site-symmetry representation at each center.
We call this set of Wyckoff positions and site-symmetry representations an {\it atomic configuration}.
Here, Wyckoff positions are taken modulo lattice translations, and the representations include the antiunitary action of magnetic symmetries.
Inducing each site-symmetry representation to the space group gives a band representation~\cite{ZakSymmetry1980,ZakBand1982,CanoBuilding2018}, which defines an occupied vector bundle equipped with crystalline symmetry actions.

However, the isomorphism class of the band representation need not determine the atomic configuration uniquely.
We address the following question:
\begin{quote}
For a Wannierizable band insulator, to what extent is the corresponding atomic configuration unique?
\end{quote}

To study the uniqueness of atomic configurations, we introduce four equivalence relations.
The four relations are defined precisely in Sec.~\ref{sec:four_equivalences}.
\begin{description}
    \item[(W) Wyckoff equivalence]
    Equivalence under continuous deformations of atomic configurations that preserve symmetry and the number of occupied orbitals.
    Wyckoff deformations include splitting and merging orbitals according to the site-symmetry representations, as well as changing representative positions by lattice translations.

    \item[(stW) Stable Wyckoff equivalence]
    Wyckoff equivalence after adding a common occupied atomic ancilla.

    \item[(ad) Adiabatic equivalence]
    The existence of a continuous deformation that preserves symmetry, the bulk gap, and the number of occupied bands in a common orbital space supplemented with unoccupied ancillary orbitals.
    The intermediate states need not be described by a continuous deformation of atomic configurations.

    \item[(st ad) Stable adiabatic equivalence]
    Adiabatic equivalence after adding a common occupied atomic ancilla.
\end{description}
Adiabatic equivalence holds precisely when the equivariant occupied bundles are isomorphic, whereas stable adiabatic equivalence corresponds to equality of their $K$-theory classes.
The four equivalence relations are related as follows:
\begin{align}
\begin{array}{ccc}
\text{Wyckoff equivalence}
&\Longrightarrow&
\text{Adiabatic equivalence}
\\[2pt]
\Downarrow&&\Downarrow
\\[2pt]
\begin{array}{c}\text{Stable Wyckoff}\\\text{equivalence}\end{array}
&\Longrightarrow&
\begin{array}{c}\text{Stable adiabatic}\\\text{equivalence}\end{array}
\end{array}.
\label{eq:rels_equ}
\end{align}
The converse implications do not hold in general, and neither adiabatic equivalence nor stable Wyckoff equivalence implies the other.
Isomorphic occupied bundles have the same topological invariants.
Thus, Wyckoff-inequivalent atomic configurations may be indistinguishable by any topological invariant of the occupied bundle.

Our main results are as follows.

First, we give an example of stable Wyckoff equivalence without adiabatic equivalence.
In a spinless $Fm'm'm'$ system, consider two configurations, each with two $s$ orbitals per primitive unit cell at one of two distinct Wyckoff positions of multiplicity one.
They are stably Wyckoff equivalent but adiabatically inequivalent with two occupied bands.
They are distinguished by the $\Z_2$ Euler class of an oriented real rank-two bundle over the boundary of the asymmetric unit, identified with the real projective plane. 

Second, we give an example of adiabatic equivalence without stable Wyckoff equivalence.
For spinless electrons in the magnetic space group $P\overline1'$, two four-orbital configurations that cannot be connected by a stable Wyckoff deformation can be connected adiabatically while keeping the number of occupied bands fixed.
A single occupied subspace therefore admits multiple Wannier bases whose atomic configurations are not stably Wyckoff equivalent.
The gauge transformation between the Wannier bases has a nontrivial three-dimensional winding number.

Third, we classify when stable Wyckoff equivalence and stable adiabatic equivalence coincide.
Stable Wyckoff equivalence always implies stable adiabatic equivalence; the question is whether the converse holds.
Stable Wyckoff and stable adiabatic equivalence coincide in two or fewer dimensions and for spinful electrons in three dimensions.
For spinless electrons in three dimensions, stable Wyckoff and stable adiabatic equivalence differ precisely when all symmetries preserve spacetime orientation and the site-symmetry group of some point contains $222$, $\overline1'$, or $\overline4'$ as a subgroup.

We now discuss the relation to previous work. 

Bradlyn et al. introduced homotopy equivalence of band representations and discussed equivalence obtained by moving orbitals through common lower-symmetry positions and inducing site-symmetry representations~\cite{BradlynTopological2017}.
This corresponds to Wyckoff equivalence up to the choice of representatives of the Wyckoff positions.
Equivariant bundle isomorphisms, however, are not restricted to homotopies of atomic configurations.
The corresponding gauge transformation need not be homotopic to the identity; large gauge transformations with nonzero three-dimensional winding number are allowed.

Huang et al. classified bosonic crystalline symmetry-protected phases by moving and merging lower-dimensional block states~\cite{HuangBuilding2017}.
This corresponds to Wyckoff equivalence for bosonic systems in our terminology. 
Real-space constructions based on gluing lower-dimensional blocks have also been developed for fermionic crystalline phases~\cite{SongRealspace2020}.

For electronic systems, Hwang et al. described stable equivalence under the motion of Wyckoff orbits and the addition of common occupied ancillas~\cite{HwangStable2026}.
The equivalence defined by Hwang et al. corresponds to stable Wyckoff equivalence in our terminology.
For all 230 nonmagnetic space groups with time-reversal symmetry, Hwang et al. tabulated stable real-space invariants as expressions in orbital multiplicities, both with and without spin-orbit coupling.

Orbital motion and the corresponding induction of site-symmetry representations are organized by the real-space Atiyah--Hirzebruch spectral sequence (AHSS)~\cite{ShiozakiGeneralized2023,ShiozakiAtiyahHirzebruch2023}.
The first differential $d^1_{1,0}$ encodes orbital motion through one-dimensional cells and the corresponding induction of representations.
Allowing formal differences of atomic configurations and quotienting by the relations encoded by $d^1_{1,0}$ gives $E^2_{0,0}$, which describes the classification under stable Wyckoff equivalence.
The $E^2$ pages for magnetic space groups have been computed systematically and are available in Ref.~\cite{ShiozakiAtiyahHirzebruch2023}.
In contrast, the higher differential
\begin{align}
    d^3_{3,-2}:E^3_{3,-2}\longrightarrow E^3_{0,0}
\end{align}
describes a pump in which creating, moving, and shrinking a two-dimensional topological phase leaves behind a zero-dimensional state~\cite{ShiozakiGeneralized2023}.
Related Chern-insulator bubble processes have also been used in real-space classifications of interacting fermionic phases with point-group symmetry~\cite{ZhangConstruction2025}.
Configuration differences in the image of $d^3_{3,-2}$ can be removed by stable adiabatic deformations, giving equivalences beyond stable Wyckoff equivalence.
Adiabatic equivalence beyond stable Wyckoff equivalence was also noted in Ref.~\cite{WadaGeneral2024} for spinless $P222$ without time-reversal symmetry.

The Chern-insulator pump itself has been studied in various contexts, including the pumping of Chern insulator wave functions to a surface~\cite{OlsenSurface2017,FidkowskiPumping2024},
surface gap closings when the surface Hamiltonian is periodic in the pump parameter~\cite{QiTopological2008,RoyTopological2011,OlsenSurface2017,LeungBulkBoundary2020},
and symmetry-enforced charge localized at point-group centers~\cite{ShiozakiGeneralized2023}.

Throughout the paper, we use the shorthand
\begin{align}
    \sum_\bk&:=\frac{\Omega}{(2\pi)^3}\int_{\mathrm{BZ}}d^3\bk,
\end{align}
where $\Omega$ is the unit-cell volume and BZ denotes the first Brillouin zone.

\section{Four equivalence relations for atomic configurations}
\label{sec:four_equivalences}

Deforming atomic configurations and deforming electronic states are different operations.
For each type of deformation, allowing or forbidding the addition of occupied ancillas gives two equivalence relations, for a total of four.
The crystalline symmetry and factor system are fixed throughout.

\subsection{Atomic configurations and equivariant occupied bundles}
\label{subsec:eq_atomic_data}

Let $G$ be a magnetic space group, $L$ the Bravais lattice, and $\Pi$ the translation subgroup.
We write the real-space action of $g\in G$ as $g\br=p_g\br+\bt_g$.
We set $\phi_g=+1$ for unitary symmetries and $\phi_g=-1$ for antiunitary symmetries, and define the factor system by
$\hat g\hat h=z_{g,h}\widehat{gh}$.
The action on momentum is then $g\bk:=\phi_gp_g\bk$.
For simplicity, we assume that lattice translations commute and take momenta modulo the reciprocal lattice.

The site-symmetry group $G_{\bfx}:=\{g\in G\mid g\bfx=\bfx\}$ fixes the position $\bfx$.
An atomic configuration $\mathcal A$ is a finite set of representative points $\bfx_i$,
one chosen from each symmetry orbit, together with the site-symmetry representations $\rho_i$ at the chosen positions:
\begin{align}
    {\cal A} = \bigoplus_{i=1}^{N_{\cal A}} (\rho_i)_{\bfx_i}.
\end{align}
Each $\rho_i$ is a direct sum of irreducible representations with nonnegative integer multiplicities.
In the presence of antiunitary symmetries, we use corepresentations for the fixed $\phi,z$
~\cite{BradleyMathematical1972}.
Here $N_{\mathcal A}$ is the number of occupied orbitals per primitive unit cell.
Changing the orbit representatives or the bases of the representations does not change the configuration.

Inducing each site-symmetry representation to $G$ and taking the direct sum over Wyckoff orbits gives the band representation
~\cite{ZakSymmetry1980,ZakBand1982,CanoBuilding2018,PoSymmetrybased2017},
\begin{align}
    \rho_{\mathcal A}
    :=\bigoplus_i\operatorname{Ind}_{G_{\bfx_i}}^G\rho_i.
    \label{eq:four_atomic_induction}
\end{align}
The Bloch representation of $\rho_{\mathcal A}$ defines the equivariant occupied bundle $E_{\mathcal A}\to\mathrm{BZ}$.
Here ``equivariant'' means that the bundle carries the action of the crystalline symmetry.
A direct sum of configurations corresponds to the direct sum of bundles,
$E_{\mathcal A\oplus\mathcal B}=E_{\mathcal A}\oplus E_{\mathcal B}$.
The bundle data do not include the spatial profiles of the Wannier orbitals.

\subsection{Wyckoff equivalence}
\label{subsec:eq_wyckoff}

\begin{dfn}\textnormal{(Wyckoff equivalence).}\enspace
\label{def:four_W}
Two atomic configurations ${\mathcal A}$ and ${\mathcal B}$ are Wyckoff equivalent,
written $\mathcal A\sim_{\mathrm W}\mathcal B$, if a finite sequence of symmetry-preserving
continuous motions of orbital centers and splittings or mergings of site-symmetry representations takes $\mathcal A$ to $\mathcal B$.
We also identify configurations that differ only by lattice translations of their site representatives,
$\bfx_i \mapsto \bfx_i + \ba, \ba \in L$.
\end{dfn}

If the site-symmetry group remains unchanged as the orbital centers move, the isomorphism class of the site-symmetry representation is preserved.
When an orbit with site-symmetry group $H$ merges at a point with a larger site-symmetry group $G_{\bfx_*}$,
the representation after merging is determined by decomposing the induced representation:
\begin{align}
    \operatorname{Ind}_{H}^{G_{\bfx_*}}\rho
    \cong\bigoplus_{\alpha}m_{\alpha}\rho_{\bfx_*}^{\alpha},
    \qquad m_{\alpha}\in\Z_{\geq0}.
    \label{eq:four_collision}
\end{align}
Splitting is the inverse operation. 

\subsubsection{A one-dimensional inversion-symmetric example}
\label{ex:one_dimensional_W}
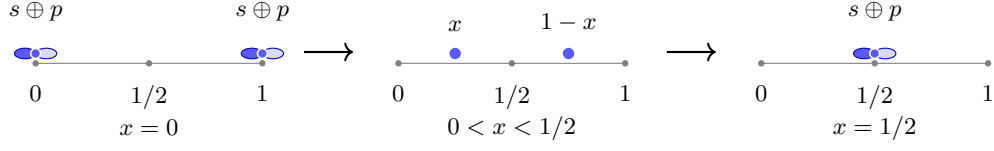
\begin{figure*}[t]
    \centering
    \begin{tikzpicture}[x=1cm,y=1cm,font=\small]
        \foreach \x in {0,4.8,9.6}{
            \draw[gray] (\x,0)--(\x+3,0);
            \foreach \d/\s in {0/0,1.5/{1/2},3/1}{
                \draw[fill=white,gray] (\x+\d,0) circle (.035);
                \node[below=5pt] at (\x+\d,0) {$\s$};
            }
        }
        \foreach \x in {0,3,11.1}{
            \filldraw[fill=blue!65,draw=blue!70!black] (\x-.14,.13) ellipse (.14 and .07);
            \filldraw[fill=blue!15,draw=blue!70!black] (\x+.14,.13) ellipse (.14 and .07);
            \filldraw[fill=blue!65,draw=white,line width=.6pt] (\x,.13) circle (.065);
            \node[above=9pt] at (\x,.13) {$s\oplus p$};
        }
        \foreach \x in {5.55,7.05}{\fill[blue!65] (\x,.13) circle (.075);}
        \node[above=5pt] at (5.55,.13) {$x$};
        \node[above=5pt] at (7.05,.13) {$1-x$};
        \draw[->,thick] (3.55,.15)--(4.2,.15);
        \draw[->,thick] (8.35,.15)--(9,.15);
        \node at (1.5,-.85) {$x=0$};
        \node at (6.3,-.85) {$0<x<1/2$};
        \node at (11.1,-.85) {$x=1/2$};
    \end{tikzpicture}
    \caption{A Wyckoff deformation in a one-dimensional inversion-symmetric system.
    The $s\oplus p$ orbitals at an inversion center split into orbitals at two symmetry-related points and merge again at the neighboring inversion center.}
    \label{fig:ex_inversion_W}
\end{figure*}

Set the lattice constant to $1$ and impose inversion $I:x\mapsto-x$.
At the inversion centers $0,1/2$, denote the even and odd one-dimensional representations by $s,p$, respectively.
Place one orbital at each point of a generic pair $\{x,-x\}$.
In both limits $x\to0$ and $x\to1/2$, the inversion-symmetric and antisymmetric combinations give $s\oplus p$.
Thus, as shown in Fig.~\ref{fig:ex_inversion_W},
\begin{align}
    s_0\oplus p_0\sim_{\mathrm W}s_{1/2}\oplus p_{1/2}.
    \label{eq:ex_inversion_W}
\end{align}
The points $x$ and $1-x$ in the figure are exchanged by inversion and merge at the same inversion center when $x=1/2$.
The points $x=0$ and $x=1$ are identified by a lattice translation.

\subsection{Stable Wyckoff equivalence}
\label{subsec:eq_stable_wyckoff}

\begin{dfn}\textnormal{(Stable Wyckoff equivalence~\cite{HwangStable2026}).}\enspace
\label{def:four_stW}
We write $\mathcal A\sim_{\mathrm{stW}}\mathcal B$
if there exists a common occupied atomic ancilla $\mathcal C$,
with finitely many orbitals per primitive unit cell, such that
\begin{align}
    \mathcal A\oplus\mathcal C
    \sim_{\mathrm W}\mathcal B\oplus\mathcal C.
    \label{eq:four_stW}
\end{align}
\end{dfn}

The stable Wyckoff equivalence classes form a commutative monoid under direct sum, with the empty configuration as the identity element.
Allowing formal differences of the stable Wyckoff equivalence classes (mathematically, group completion) gives the abelian group $E^2_{0,0}$ of the real-space AHSS~\cite{ShiozakiGeneralized2023,ShiozakiAtiyahHirzebruch2023}.

We give two examples of configurations that are stably Wyckoff equivalent but not Wyckoff equivalent.

\subsubsection{Two-dimensional $pmm$: exchanging site-symmetry representations at fixed positions}
\label{ex:pmm_stW}
\begin{figure*}[t]
    \centering
    \begin{tikzpicture}[x=1cm,y=1cm,font=\small]
        \foreach \o in {0,7}{
            \draw[gray] (\o,0) rectangle (\o+2.6,2.6);
            \foreach \x/\y/\s in {0/0/a,2.6/0/b,2.6/2.6/c,0/2.6/d}{
                \node[below=8pt] at (\o+\x,\y) {$\s$};
            }
        }
        \foreach \x/\y/\r in {0/0/0,2.6/0/90,2.6/2.6/0,0/2.6/90,
                                  7/0/90,9.6/0/0,9.6/2.6/90,7/2.6/0}{
            \begin{scope}[shift={(\x,\y)},rotate=\r]
                \filldraw[fill=blue!65,draw=blue!70!black] (-.17,0) ellipse (.16 and .075);
                \filldraw[fill=blue!15,draw=blue!70!black] (.17,0) ellipse (.16 and .075);
            \end{scope}
        }
        \foreach \x in {0,7}{\filldraw[fill=green!55!black,draw=white,line width=.8pt] (\x,0) circle (.09);}
        \draw[->,thick,green!40!black] (.6,.2)--(2,.2);
        \draw[->,thick,green!40!black] (2.4,.6)--(2.4,2);
        \draw[->,thick,green!40!black] (2,2.4)--(.6,2.4);
        \draw[->,thick,green!40!black] (.2,2)--(.2,.6);
        \draw[->,thick] (3.5,1.3)--(6.1,1.3);
        \node[above=4pt] at (4.8,1.3) {$\sim_{\mathrm W}$};
        \node at (1.3,-1) {$\mathcal A\oplus s_a$};
        \node at (8.3,-1) {$\mathcal B\oplus s_a$};
    \end{tikzpicture}
    \caption{Stable Wyckoff equivalence in $pmm$.
    Horizontal and vertical dumbbells denote $p_x$ and $p_y$ orbitals, respectively.
    The green circle is the common occupied ancilla $s_a$.
    The orbital pairs in Eq.~\eqref{eq:ex_pmm_cycle} move in the order indicated by the arrows.}
    \label{fig:ex_pmm_stW}
\end{figure*}
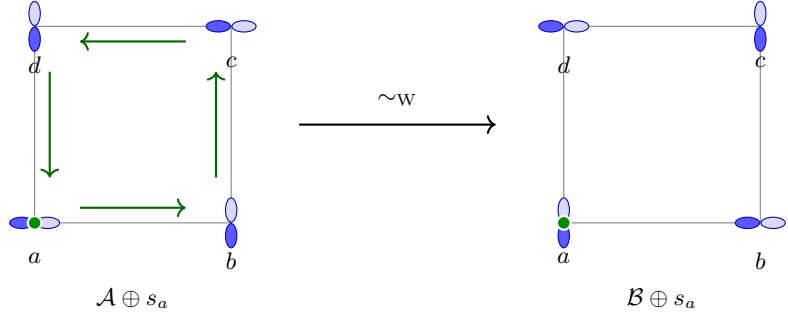

Consider a spinless system on a primitive rectangular lattice with mirror symmetries
$m_x:(x,y)\mapsto(-x,y)$ and $m_y:(x,y)\mapsto(x,-y)$.
Label the site-symmetry representations by the eigenvalues of $m_x,m_y$ as
\begin{align}
    s=(+,+),\qquad p_x=(-,+),\qquad p_y=(+,-).
\end{align}
At the four high-symmetry points
$a=(0,0),b=(1/2,0),c=(1/2,1/2),d=(0,1/2)$,
consider the two configurations in Fig.~\ref{fig:ex_pmm_stW},
\begin{align}
    &\mathcal A = (p_x)_a \oplus (p_y)_b\oplus (p_x)_c\oplus (p_y)_d, \notag\\
    &\mathcal B = (p_y)_a \oplus (p_x)_b\oplus (p_y)_c\oplus (p_x)_d.
    \label{eq:ex_pmm_configs}
\end{align}
There is only one orbital at each intersection of mirror lines, whereas moving away from an intersection requires an orbit of at least two points.
Each orbital is therefore pinned to its initial high-symmetry position, and $\mathcal A\not\sim_{\mathrm W}\mathcal B$.

Adding one occupied $s$ orbital at the Wyckoff position $a$ as an ancilla allows pairs of orbitals to move in the following order:
\begin{align}
    &(s \oplus p_x)_a
    \sim_{\rm W} (s \oplus p_x)_b, \quad
    (s \oplus p_y)_b
    \sim_{\rm W} (s \oplus p_y)_c, \notag\\
    &(s \oplus p_x)_c
    \sim_{\rm W} (s \oplus p_x)_d, \quad
    (s \oplus p_y)_d
    \sim_{\rm W} (s \oplus p_y)_a.
    \label{eq:ex_pmm_cycle}
\end{align}
For example, two orbitals even under $m_y$ can move horizontally and decompose into $s\oplus p_x$ at either endpoint.
Similarly, $s\oplus p_y$ can move vertically.
After one cycle, $p_x$ and $p_y$ are exchanged at each point, and the $s$ orbital returns to $a$. Hence
\begin{align}
    \mathcal A\oplus s_a\sim_{\mathrm W}\mathcal B\oplus s_a,
    \qquad \mathcal A\sim_{\mathrm{stW}}\mathcal B.
    \label{eq:ex_pmm_stW}
\end{align}

\subsubsection{Three-dimensional $P222$: changing the occupied positions}
\label{ex:P222_stW}
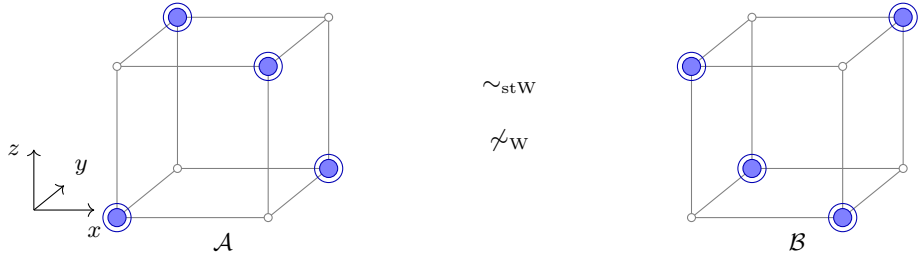
\begin{figure*}[t]
    \centering
    \begin{tikzpicture}[x={(2cm,0cm)},y={(.8cm,.65cm)},z={(0cm,2cm)},font=\small]
        \foreach \o/\exparity/\name in {0/0/{\mathcal A},3.8/1/{\mathcal B}}{
            \begin{scope}[shift={(\o,0,0)}]
                \draw[gray] (0,0,0)--(1,0,0)--(1,1,0)--(0,1,0)--cycle;
                \draw[gray] (0,0,1)--(1,0,1)--(1,1,1)--(0,1,1)--cycle;
                \foreach \x/\y in {0/0,1/0,1/1,0/1}{\draw[gray] (\x,\y,0)--(\x,\y,1);}
                \foreach \x in {0,1}{\foreach \y in {0,1}{\foreach \z in {0,1}{
                    \pgfmathtruncatemacro{\pmod}{mod(\x+\y+\z,2)}
                    \ifnum\pmod=\exparity
                        \filldraw[fill=blue!50,draw=blue!70!black] (\x,\y,\z) circle (3.4pt);
                        \draw[blue!70!black] (\x,\y,\z) circle (5.3pt);
                    \else
                        \filldraw[fill=white,draw=gray] (\x,\y,\z) circle (1.5pt);
                    \fi
                }}}
                \node[below=12pt] at (.5,.5,0) {$\name$};
            \end{scope}
        }
        \node at (2.4,.5,.7) {$\sim_{\mathrm{stW}}$};
        \node at (2.4,.5,.35) {$\not\sim_{\mathrm W}$};
        \draw[->] (-.55,0,.05)--(-.15,0,.05) node[below=3pt] {$x$};
        \draw[->] (-.55,0,.05)--(-.55,.5,.05) node[above right=1pt] {$y$};
        \draw[->] (-.55,0,.05)--(-.55,0,.45) node[left=2pt] {$z$};
    \end{tikzpicture}
    \caption{Stable Wyckoff equivalence in $P222$.
    The eight vertices of the cube are the half-lattice points $\br_\epsilon=\epsilon/2$,
    and each edge has a length equal to half the corresponding lattice constant. Double circles denote $2s$.
    The left and right configurations occupy the even and odd vertices, respectively.
    After adding the four occupied ancillary orbitals in Eq.~\eqref{eq:ex_P222_ancilla},
    $\mathcal A$ and $\mathcal B$ can be connected by the path in Eq.~\eqref{eq:ex_P222_moves}.}
    \label{fig:ex_P222_stW}
\end{figure*}

Consider a spinless system on a primitive orthorhombic lattice with twofold rotations about each coordinate axis.
The four one-dimensional representations of the site-symmetry group $222$ are
\begin{align}
    \begin{array}{c|ccc}
        &C_{2x}&C_{2y}&C_{2z}\\\hline
        s&+1&+1&+1\\
        p_x&+1&-1&-1\\
        p_y&-1&+1&-1\\
        p_z&-1&-1&+1
    \end{array}.
\end{align}
For the half-lattice points $\br_\epsilon=(\epsilon_x,\epsilon_y,\epsilon_z)/2$,
$\epsilon_j\in\{0,1\}$, write $|\epsilon|=\epsilon_x+\epsilon_y+\epsilon_z$.
The two configurations in Fig.~\ref{fig:ex_P222_stW},
\begin{align}
    \mathcal A=\bigoplus_{|\epsilon|\,\mathrm{even}}2s_{\br_\epsilon},
    \qquad
    \mathcal B=\bigoplus_{|\epsilon|\,\mathrm{odd}}2s_{\br_\epsilon},
    \label{eq:ex_P222_configs}
\end{align}
each have two orbitals at each of four points, giving eight occupied orbitals in total.

A two-point orbit along a twofold axis gives a sum of two distinct irreducible representations when it merges at a high-symmetry endpoint.
Along the $x$ axis, for example, the sum is $s\oplus p_x$ or $p_y\oplus p_z$, but never $2s$.
Splitting into generic positions requires four orbitals.
Thus, the $2s$ orbitals at each point cannot move away from that point, and $\mathcal A\not\sim_{\mathrm W}\mathcal B$.

Choose the common occupied ancilla
\begin{align}
    \mathcal C=(p_x\oplus p_y)_{001}\oplus(p_x\oplus p_y)_{111}.
    \label{eq:ex_P222_ancilla}
\end{align}
Here a subscript such as $001$ denotes $\br_{001}=(0,0,1/2)$.
The pairs $s\oplus p_x,s\oplus p_y,p_x\oplus p_y$ can move in the $x,y,z$ directions, respectively.
We can therefore perform the following Wyckoff deformation in four steps:
\begin{align}
    \begin{array}{c|l}
        1&z:001\to000,\quad z:111\to110\\
        2&x:000\to100,\quad y:000\to010,\\
         &x:110\to010,\quad y:110\to100\\
        3&z:100\to101,\quad z:010\to011\\
        4&x:101\to001,\quad y:101\to111,\\
         &x:011\to111,\quad y:011\to001
    \end{array}.
    \label{eq:ex_P222_moves}
\end{align}
Step 1 moves $p_x,p_y$ to $000,110$. Step 2 sends the two $s$ orbitals at each point along different edges.
Step 3 moves $p_x,p_y$ while leaving $2s$ at $100,010$. Step 4 repeats the same operation.
The occupied orbitals of $\mathcal C$ return to their initial positions at the end. Thus
\begin{align}
    \mathcal A\oplus\mathcal C\sim_{\mathrm W}\mathcal B\oplus\mathcal C,
    \qquad \mathcal A\sim_{\mathrm{stW}}\mathcal B.
\end{align}
In the $pmm$ example, the number of orbitals at each occupied position was unchanged.
Here the occupied positions themselves change.

\subsection{Adiabatic equivalence of atomic configurations}
\label{subsec:eq_atomic_adiabatic}

Section~\ref{subsec:eq_stable_wyckoff} defined equivalence through continuous deformations of atomic configurations.
Adiabatic deformations of band insulators are more general: they can mix occupied states with unoccupied degrees of freedom.

Consider two atomic configurations with the same number of occupied orbitals, $N_{\mathcal A}=N_{\mathcal B}=N$.
Fix representative orbitals in the unit cell, and let $\Phi_a(\bk)$ be the periodic orthonormal basis formed by the atomic orbitals of configuration $a=\mathcal A,\mathcal B$.
The symmetry matrices $u^a_{\bk}(g)$ give the Bloch form of the band representation determined by the atomic configuration.
They are defined by their action on the coefficient vector $\psi \in\C^N$:
\begin{align}
    \hat g\bigl(\Phi_a(\bk)\psi\bigr)
    =\Phi_a(g\bk)u^a_{\bk}(g)\psi^{\phi_g}.
    \label{eq:atomic_ad_sewing}
\end{align}
Here $M^{\phi_g}=M$ for unitary symmetries and $M^{\phi_g}=M^*$ for antiunitary symmetries.
For explicit constructions of $u_\bk^a(g)$, see Refs.~\cite{CanoBuilding2018,PoSymmetrybased2017}.
The symmetry actions on the two configurations obey the same factor system.

Treat the orbitals of the two configurations as independent and use the single-particle Hilbert space
$\mathcal H_{\bk}=\C^N_{\mathcal A}\oplus\C^N_{\mathcal B}$ at each momentum.
Order the blocks as written and fix the symmetry matrices to
$u_{\bk}(g):=u^{\mathcal A}_{\bk}(g)\oplus u^{\mathcal B}_{\bk}(g)$.
The atomic limits with $\mathcal A$ occupied and $\mathcal B$ empty, and with $\mathcal A$ empty and $\mathcal B$ occupied, are represented by
\begin{align}
    H_0(\bk)=\begin{pmatrix}-I_N&0\\0&I_N\end{pmatrix},
    \qquad
    H_1(\bk)=\begin{pmatrix}I_N&0\\0&-I_N\end{pmatrix},
    \label{eq:atomic_ad_endpoints}
\end{align}
respectively, with the Fermi energy set to zero.
At each endpoint, the other configuration supplies unoccupied ancillary orbitals, and the number of occupied orbitals is $N$.
Adiabatic equivalence allows unoccupied ancillary orbitals but does not add occupied ancillas.

\begin{dfn}\textnormal{(Adiabatic equivalence of atomic configurations).}\enspace
\label{def:atomic_ad}
We write $\mathcal A\sim_{\mathrm{ad}}\mathcal B$ if a family of Hermitian matrices $H_t(\bk)$ connects the endpoints in Eq.~\eqref{eq:atomic_ad_endpoints},
is continuous in $(\bk,t)$ and periodic in momentum, remains gapped at zero energy for all $(\bk,t)$, and satisfies
\begin{align}
    u_{\bk}(g)H_t(\bk)^{\phi_g}
    =H_t(g\bk)u_{\bk}(g),
    \qquad g\in G.
    \label{eq:atomic_ad_symmetry}
\end{align}
\end{dfn}

The adiabatic deformation $H_t(\bk)$ allows mixing between atomic orbitals.
The change in the occupied states need not be realized by continuous motion of orbital centers.

\begin{prop}\textnormal{(Criterion for adiabatic equivalence of atomic configurations).}\enspace
\label{prop:atomic_ad_intertwiner}
The relation $\mathcal A\sim_{\mathrm{ad}}\mathcal B$ holds if and only if there exists a continuous, periodic unitary matrix $V_{\bk}\in U(N)$ satisfying
\begin{align}
    u^{\mathcal A}_{\bk}(g)V_{\bk}^{\phi_g}
    =V_{g\bk}u^{\mathcal B}_{\bk}(g),
    \qquad g\in G.
    \label{eq:atomic_ad_intertwiner}
\end{align}
\end{prop}

\noindent\textit{Proof.} 
Given a unitary $V_{\bk}$ satisfying Eq.~\eqref{eq:atomic_ad_intertwiner}, set
\begin{align}
    H_t(\bk)=
    \begin{pmatrix}
        -\cos(\pi t)I_N&\sin(\pi t)V_{\bk}\\
        \sin(\pi t)V_{\bk}^{\dagger}&\cos(\pi t)I_N
    \end{pmatrix}.
    \label{eq:atomic_ad_path}
\end{align}
Conversely, suppose that an adiabatic path $H_t(\bk)$ exists.
We may choose $H_t(\bk)$ to be smooth in $t$ while preserving the endpoint Hamiltonians and symmetry.
Flatten the Hamiltonian as $Q_t(\bk) = H_t(\bk)[H_t(\bk)^2]^{-1/2}$ and define the occupied projector by $P_t(\bk) = (I_{2N}-Q_t(\bk))/2$.
Let $U_t(\bk)$ be the unique solution of the parallel-transport equation~\cite{KatoAdiabatic1950},
\begin{align}
    \p_t U_t(\bk)
    &=[\p_tP_t(\bk),P_t(\bk)]U_t(\bk),
    &U_0(\bk)&=I_{2N}.
    \label{eq:atomic_ad_transport_ode}
\end{align}
The generator on the right-hand side is anti-Hermitian, so $U_t(\bk)$ is unitary.
Differentiating $P_t^2=P_t$ gives $[[\p_tP_t,P_t],P_t]=\p_tP_t$.
Hence $\p_t\bigl(U_t(\bk)^\dagger P_t(\bk)U_t(\bk)\bigr)=0$, and
\begin{align}
    P_t(\bk)=U_t(\bk)P_0(\bk)U_t(\bk)^\dagger.
    \label{eq:atomic_ad_transport_projector}
\end{align}
Continuous dependence on parameters and uniqueness of the solution imply that $U_t(\bk)$ is continuous and periodic in momentum.
The symmetry matrices $u_{\bk}(g)$ are independent of $t$.
Equation~\eqref{eq:atomic_ad_symmetry} then implies that the projector and the generator obey the same symmetry transformation law.
Uniqueness of the solution therefore gives
\begin{align}
    u_{\bk}(g)U_t(\bk)^{\phi_g}
    =U_t(g\bk)u_{\bk}(g).
    \label{eq:atomic_ad_transport_symmetry}
\end{align}
Since $P_0 = {\rm diag}(I_N,O)$ and $P_1={\rm diag}(O,I_N)$, we have
\begin{align}
    U_1(\bk) = \begin{pmatrix}
        O & V_\bk\\
        W_\bk& O\\
    \end{pmatrix}.
\end{align}
The block $V_{\bk}\in U(N)$ is continuous and periodic and satisfies Eq.~\eqref{eq:atomic_ad_intertwiner}.
\hfill$\square$\par

Although $U_t(\bk)$ itself is continuously connected to $I_{2N}$,
the off-diagonal block $V_{\bk}$ need not be continuously deformable to the identity.

We next consider how Wannier states evolve during an adiabatic deformation.
To discuss localization, assume that $H_t(\bk)$ is periodic in momentum and extends to a matrix family
that is smooth in $t$ and analytic in momentum on a complex neighborhood of the BZ independent of $t$.
Using the solution of Eq.~\eqref{eq:atomic_ad_transport_ode}, choose the occupied frame as
\begin{align}
    \Phi_0(\bk)=\begin{pmatrix}I_N\\O\end{pmatrix},
    \qquad
    \Phi_t(\bk)=U_t(\bk)\Phi_0(\bk).
    \label{eq:atomic_ad_wannier_frame}
\end{align}
The frame $\Phi_t(\bk)$ is smooth in $t$ and periodic and analytic in momentum.
Equation~\eqref{eq:atomic_ad_transport_symmetry} also gives
\begin{align}
    u_{\bk}(g)\Phi_t(\bk)^{\phi_g}
    =\Phi_t(g\bk)u^{\mathcal A}_{\bk}(g),\quad 
    t \in [0,1], 
    \label{eq:atomic_ad_wannier_sewing}
\end{align}
so the symmetry transformation law of the initial atomic configuration $\mathcal A$ is preserved.

Write the atomic orbitals of configuration $\mathcal A\oplus\mathcal B$ as
$\ket{\bR+\bw_i}$, $i=1,\ldots,2N$.
The index $i$ distinguishes orbitals even at the same position.
The first $N$ orbitals correspond to $\mathcal A$, and the rest to $\mathcal B$.
The Wannier states defined by Fourier transforming the frame $\Phi_t(\bk)$ satisfy
\begin{align}
    \Braket{\bR'+\bw_i|W_{n\bR}(t)}
    =\sum_\bk[\Phi_t(\bk)]_{in}
      e^{-i\bk\cdot(\bR-\bR')}.
    \label{eq:atomic_ad_wannier_components}
\end{align}
Under the analyticity and gap assumptions above, the Wannier states $\ket{W_{n\bR}(t)}$ are exponentially localized~\cite{PanatiTriviality2007,BrouderExponential2007}.
The wave functions and Wannier centers of $\ket{W_{n\bR}(t)}$,
\begin{align}
    \br_n(t)
    :=\bra{W_{n\mathbf 0}(t)}\hat{\br}\ket{W_{n\mathbf 0}(t)},
    \label{eq:atomic_ad_wannier_centers}
\end{align}
vary continuously with $t$.

The centers and site-symmetry representations of the Wannier basis $\ket{W_{n\bR}(t)}$ determine an atomic configuration,
which varies continuously in the sense of Sec.~\ref{subsec:eq_wyckoff}.
The deformation of the atomic configuration defined by $\ket{W_{n\bR}(t)}$ starts from $\mathcal A$, but need not end at $\mathcal B$.
The fact that the Wannier states lie in $\mathcal H_{\mathcal B}$ at $t=1$ means only that
{\it the Wannier states can be expanded using the atomic orbitals of $\mathcal B$ alone.}
The Wannier states and the atomic orbitals used in the expansion need not have the same centers or site-symmetry representations.

Indeed, at $t=1$ the occupied subspace equals $\mathcal H_{\mathcal B}$, but the atomic orbital basis of configuration $\mathcal B$ is given by the frame
\begin{align}
    \Psi_1(\bk)=\begin{pmatrix}O\\I_N\end{pmatrix}.
\end{align}
In terms of the lower-left block of $U_1(\bk)$ defined above, the transported frame satisfies
\begin{align}
    \Phi_1(\bk)
    =\begin{pmatrix}O\\W_\bk\end{pmatrix}
    =\Psi_1(\bk)W_\bk,\notag\\
    u^{\mathcal B}_{\bk}(g)W_\bk^{\phi_g}
    =W_{g\bk}u^{\mathcal A}_{\bk}(g).
    \label{eq:atomic_ad_wannier_endpoint}
\end{align}
The gauge transformation between $\Psi_1(\bk)$ and $\Phi_1(\bk)$ is therefore $W_\bk$.
Writing the atomic orbitals of $\mathcal B$ as $\ket{\bR'+\bw_j^{\mathcal B}}$, $j=1,\ldots,N$,
the Wannier states at the endpoint have the expansion
\begin{align}
    \ket{W_{n\bR}(1)}
    =\sum_{\bR'}\sum_{j=1}^{N}
      \ket{\bR'+\bw_j^{\mathcal B}}
      \sum_\bk[W_\bk]_{jn}e^{-i\bk\cdot(\bR-\bR')}.
    \label{eq:atomic_ad_wannier_endpoint_fourier}
\end{align}
The Fourier coefficients decay exponentially with $|\bR-\bR'|$, but are generally nonzero for atomic orbitals in multiple unit cells.
For example, in a lattice model whose atomic orbitals form an eigenbasis of the position operator, the Wannier center is
\begin{align}
    \br_n(1)
    =\sum_{\bR'}\sum_{j=1}^{N}
      (\bR'+\bw_j^{\mathcal B})
      \left|\sum_\bk[W_\bk]_{jn}e^{i\bk\cdot\bR'}\right|^2.
    \label{eq:atomic_ad_wannier_endpoint_center}
\end{align}
The Wannier center $\br_n(1)$ need not coincide with any atomic orbital position appearing in the expansion.
Similarly, the site-symmetry representation must be read from the symmetry action on the set of Wannier states,
rather than from the representations of the individual atomic orbitals in the expansion.
Thus, even though all Wannier states lie in $\mathcal H_{\mathcal B}$ at the endpoint,
their centers and site-symmetry representations need not define the atomic configuration $\mathcal B$.
We therefore cannot infer $\mathcal A\sim_{\mathrm W}\mathcal B$ from the transport of the Wannier basis by $U_t(\bk)$.
Section~\ref{sec:adia_wannier_center} gives an explicit example of Wannier bases that define different atomic configurations within the same occupied subspace.

\subsection{Stable adiabatic equivalence}
\label{subsec:eq_stable_adiabatic}

\begin{dfn}\textnormal{(Stable adiabatic equivalence).}\enspace
\label{def:four_stad}
We write $\mathcal A\sim_{\mathrm{st\,ad}}\mathcal B$ if a common finite atomic configuration $\mathcal C$
with the same symmetry and factor system can be added as an occupied ancilla so that
\begin{align}
    \mathcal A\oplus\mathcal C
    \sim_{\mathrm{ad}}\mathcal B\oplus\mathcal C.
    \label{eq:four_stad}
\end{align}
\end{dfn}

Here $\sim_{\mathrm{ad}}$ denotes adiabatic equivalence of atomic configurations as defined in Definition~\ref{def:atomic_ad}.

Stable adiabatic equivalence is equivalent to $[E_{\mathcal A}]=[E_{\mathcal B}]$ in the $K$-group~\cite{FreedTwisted2013}.
Indeed, equality in the $K$-group gives an isomorphism of bundles
$E_{\mathcal A}\oplus F\cong_G E_{\mathcal B}\oplus F$ for some occupied bundle $F$, which need not be Wannierizable.
Let $E_{\mathcal C}$ be a full atomic orbital bundle containing $F$, and let $F^\perp$ be the orthogonal complement of $F$ in $E_{\mathcal C}$.
Then $F\oplus F^\perp\cong E_{\mathcal C}$.
Adding $F^\perp$ to both sides and applying Proposition~\ref{prop:atomic_ad_intertwiner} gives
$\mathcal A\oplus\mathcal C\sim_{\mathrm{ad}}\mathcal B\oplus\mathcal C$.

\subsection{Comparison of the four equivalence relations}
\label{subsec:eq_comparison}

A Wyckoff deformation continuously deforms the band representation, so Wyckoff equivalence implies adiabatic equivalence.
Applying the same argument after adding a common atomic ancilla shows that stable Wyckoff equivalence implies stable adiabatic equivalence.
Each stable relation also includes the case without an ancilla.
Together, these observations establish all the implications in Eq.~\eqref{eq:rels_equ}.

We next illustrate the differences through explicit examples and then turn to the classification using the real-space AHSS.

\section{Spinless \texorpdfstring{$Fm'm'm'$}{Fm'm'm'}: stable Wyckoff equivalence without adiabatic equivalence}
\label{sec:fcc:example}

The pairs of atomic configurations in the $pmm$ and $P222$ examples in Secs.~\ref{ex:pmm_stW} and \ref{ex:P222_stW} are both stably Wyckoff equivalent and adiabatically equivalent without adding occupied bands.
Appendix~\ref{app:examples_intertwiners} gives explicit finite-range models.
Here we give an example of stable Wyckoff equivalence without adiabatic equivalence at a fixed number of occupied bands.
The two atomic configurations considered here become Wyckoff equivalent after adding a common occupied ancilla, but the corresponding occupied bundles are not isomorphic.
We also discuss a topological invariant that distinguishes the two occupied bundles.

\subsection{Lattice and atomic configurations}
\label{subsec:fcc:configurations}

We set the side length of the conventional unit cell to $1$ and define the face-centered lattice $L$ and the dual lattice $L^*$ by
\begin{align}
 L&=\left\{\tfrac12(n_x,n_y,n_z):\begin{array}{l}n_i\in\Z,\\n_x+n_y+n_z\in2\Z\end{array}\right\},
 \label{eq:fcc:lattice}\\
 L^*&=\{\bm m\in\Z^3:m_x\equiv m_y\equiv m_z\pmod2\},\notag\\
 &\qquad \mathrm{BZ}=\R^3/(2\pi L^*).
 \label{eq:fcc:reciprocal}
\end{align}
We write $\bm e_i$ for the unit vectors along the coordinate axes.
In addition to lattice translations, we impose the unitary twofold rotations
\begin{align}
 R_x&=\operatorname{diag}(1,-1,-1),\notag\\
 R_y&=\operatorname{diag}(-1,1,-1),\quad
 R_z=R_xR_y,
 \label{eq:fcc:rotations}
\end{align}
and the antiunitary PT operation $h=PT$.
The operation $h$ acts as inversion in real space and as the identity in momentum space,
with $h^2=1$ and $hR_i=R_i h$.
Lattice translations, the twofold rotations $R_i$, and $PT$ generate the magnetic space group $Fm'm'm'$.
We take the factor system to be trivial, corresponding to a spinless system.
Neither spatial inversion $P$ nor time reversal $T$ is assumed to be a symmetry on its own.

Consider the two Wyckoff positions
\begin{align}
 a&=(0,0,0),\notag\\ b&=\tfrac12(1,1,1)
 \equiv q=\tfrac12(1,0,0)\pmod L.
 \label{eq:fcc:sites}
\end{align}
Although $a$ and $b$ are not related by a lattice translation,
$\bm e_x/2$, $\bm e_y/2$, and $\bm e_z/2$ are representatives of the same position $b$.
Below, we use the representative $q$ to simplify the calculation.
The site-symmetry group at $r=a,q$ is generated by
\begin{align}
 g_i^{(r)}=\{R_i\mid r-R_ir\},\qquad
 h^{(r)}=\{-I\mid2r\}T.
 \label{eq:fcc:site_operations}
\end{align}
The one-dimensional representations of the site-symmetry group at $r$ are listed below:
\begin{center}
\begin{tabular}{c|ccc|c}
 & $g_x^{(r)}$ & $g_y^{(r)}$ & $g_z^{(r)}$ & $h^{(r)}$ \\ \hline
 $s$   & $+1$ & $+1$ & $+1$ & $K$ \\
 $p_x$ & $+1$ & $-1$ & $-1$ & $K$ \\
 $p_y$ & $-1$ & $+1$ & $-1$ & $K$ \\
 $p_z$ & $-1$ & $-1$ & $+1$ & $K$
\end{tabular}
\end{center}
Here, $K$ denotes complex conjugation of the coefficients.

The atomic configurations to be compared and the common occupied ancilla are
\begin{align}
 \mathcal A=2s_a,\qquad \mathcal B=2s_b,\qquad
 \mathcal C=(p_x\oplus p_y)_a.
 \label{eq:fcc:ABC}
\end{align}
Both $\mathcal A$ and $\mathcal B$ contain two occupied orbitals per primitive unit cell.

\begin{figure*}[t]
\centering
\begin{tikzpicture}[font=\small]
\begin{scope}[x={(2.7cm,0cm)},y={(0.95cm,0.65cm)},z={(0cm,2.7cm)}]
 \foreach \z in {0,1}{
  \draw[gray!65] (0,0,\z)--(1,0,\z)--(1,1,\z)--(0,1,\z)--cycle;
 }
 \foreach \x in {0,1}{\foreach \y in {0,1}{\draw[gray!65] (\x,\y,0)--(\x,\y,1);}}
 \foreach \x in {0,1}{\foreach \y in {0,1}{\foreach \z in {0,1}{
  \fill[blue!65!black] (\x,\y,\z) circle (1.8pt);
 }}}
 \foreach \p in {(0.5,0.5,0),(0.5,0.5,1),(0.5,0,0.5),(0.5,1,0.5),(0,0.5,0.5),(1,0.5,0.5)}{
  \fill[blue!65!black] \p circle (1.8pt);
 }
 \draw[densely dashed,gray] (0.5,0.5,0)--(0.5,0.5,1);
 \node[below left] at (0,0,0) {$a$};
 \node at (0.5,0.5,-0.27) {$\mathcal A=2s_a$};
\end{scope}
\begin{scope}[shift={(6.6cm,0cm)},x={(2.7cm,0cm)},y={(0.95cm,0.65cm)},z={(0cm,2.7cm)}]
 \foreach \z in {0,1}{
  \draw[gray!65] (0,0,\z)--(1,0,\z)--(1,1,\z)--(0,1,\z)--cycle;
 }
 \foreach \x in {0,1}{\foreach \y in {0,1}{\draw[gray!65] (\x,\y,0)--(\x,\y,1);}}
 \foreach \x in {0,1}{\foreach \y in {0,1}{\foreach \z in {0,1}{
  \fill[gray!50] (\x,\y,\z) circle (1pt);
 }}}
 \foreach \i in {0,1}{\foreach \j in {0,1}{
  \fill[blue!65!black] (0.5,\i,\j) circle (1.8pt);
  \fill[blue!65!black] (\i,0.5,\j) circle (1.8pt);
  \fill[blue!65!black] (\i,\j,0.5) circle (1.8pt);
 }}
 \fill[blue!65!black] (0.5,0.5,0.5) circle (2.4pt);
 \node[above right] at (0.5,0.5,0.5) {$b$};
 \node[below left=2pt] at (0.5,0,0) {$q$};
 \node at (0.5,0.5,-0.27) {$\mathcal B=2s_b$};
\end{scope}
\node[align=center] at (5.15cm,1.6cm)
 {$\sim_{\mathrm{stW}}$\\[6pt]$\not\sim_{\mathrm{ad}}$};
\end{tikzpicture}
\caption{Two atomic configurations in spinless $Fm'm'm'$.
Each blue point represents two occupied $s$ orbitals.
The corners and face centers on the left are related by lattice translations, as are the body center and edge midpoints on the right.
The cube is a conventional unit cell with side length $1$. Each configuration has two occupied orbitals per primitive unit cell.}
\label{fig:fcc:configurations}
\end{figure*}
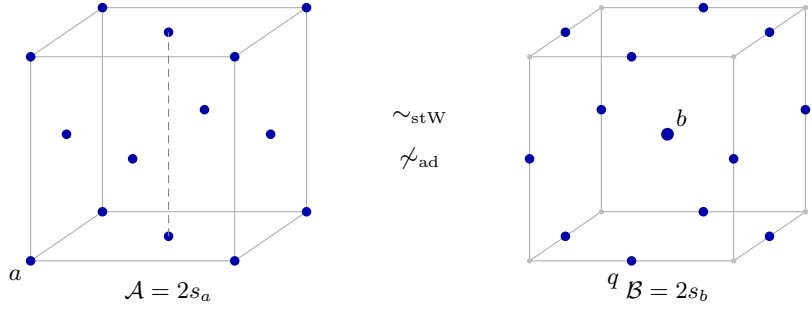

\begin{prop}
\label{prop:fcc:main}
The configurations in Eq.~\eqref{eq:fcc:ABC} satisfy
\begin{align}
 \mathcal A\sim_{\mathrm{stW}}\mathcal B,
 \qquad \mathcal A\not\sim_{\mathrm{ad}}\mathcal B.
 \label{eq:fcc:claim}
\end{align}
\end{prop}

\subsection{Wyckoff deformation with an occupied ancilla}
\label{subsec:fcc:stable_path}
As in Sec.~\ref{ex:P222_stW}, the $s\oplus p_x$, $s \oplus p_y$, and $p_x \oplus p_y$ orbitals can move continuously along the $x$, $y$, and $z$ directions, respectively.
For $0\leq u\leq1$, we perform the following three operations:
\begin{enumerate}
 \item Split $(s\oplus p_x)_a$ into orbitals at $\pm u\bm e_x/2$
 and merge the resulting pair into $(s\oplus p_x)_b$ at the endpoint.
 \item Split the remaining $(s\oplus p_y)_a$ into orbitals at $\pm u\bm e_y/2$
 and merge the resulting pair into $(s\oplus p_y)_b$ at the endpoint.
 \item Split $(p_x\oplus p_y)_b$ into orbitals at $q\pm u\bm e_z/2$
 and merge the resulting pair into $(p_x\oplus p_y)_a$ at the endpoint.
\end{enumerate}
Thus,
\begin{align}
 2s_a\oplus(p_x)_a\oplus(p_y)_a
 &\sim_{\mathrm W}s_a\oplus s_b\oplus(p_x)_b\oplus(p_y)_a\notag\\
 &\sim_{\mathrm W}2s_b\oplus(p_x)_b\oplus(p_y)_b\notag\\
 &\sim_{\mathrm W}2s_b\oplus(p_x)_a\oplus(p_y)_a.
 \label{eq:fcc:positive_path}
\end{align}
Since the same ancilla $\mathcal C$ appears at both endpoints, this deformation establishes $\mathcal A\sim_{\mathrm{stW}}\mathcal B$.

\subsection{Proof of adiabatic inequivalence}
\label{subsec:fcc:no_isomorphism}

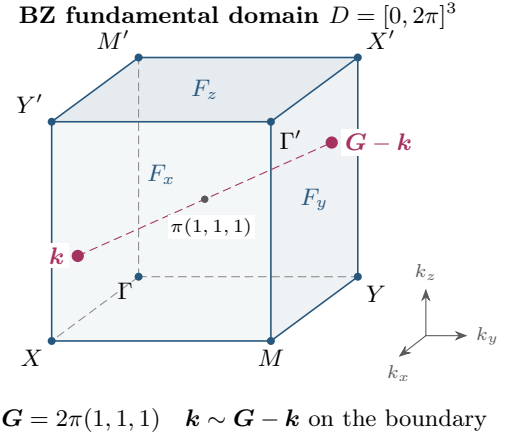
\begin{figure}[tbp]
\centering
\begin{tikzpicture}[font=\small,>=Stealth,line join=round,line cap=round]
  \definecolor{fccblue}{RGB}{40,87,125}
  \definecolor{fccpair}{RGB}{165,49,95}
  \node[anchor=west,font=\small\bfseries] at (0,4.75)
    {BZ fundamental domain $D=[0,2\pi]^3$};
  % Coordinates are k/(2 pi); x points down-left, y right, z up.
  \begin{scope}[shift={(1.7,1.3)},x={(-1.15cm,-0.85cm)},
      y={(2.9cm,0cm)},z={(0cm,2.9cm)}]
    \coordinate (G) at (0,0,0);
    \coordinate (X) at (1,0,0);
    \coordinate (Y) at (0,1,0);
    \coordinate (M) at (1,1,0);
    \coordinate (Gp) at (1,1,1);
    \coordinate (Xp) at (0,1,1);
    \coordinate (Yp) at (1,0,1);
    \coordinate (Mp) at (0,0,1);
    \coordinate (O) at (.5,.5,.5);
    \coordinate (P) at (.70,0,.30);
    \coordinate (Q) at (.30,1,.70);
    \fill[fccblue!5] (X)--(M)--(Gp)--(Yp)--cycle;
    \fill[fccblue!8] (M)--(Y)--(Xp)--(Gp)--cycle;
    \fill[fccblue!12] (Yp)--(Gp)--(Xp)--(Mp)--cycle;
    \node[fccblue] at (1,.5,.76) {$F_x$};
    \node[fccblue] at (.5,1,.5) {$F_y$};
    \node[fccblue] at (.5,.5,1) {$F_z$};
    \draw[densely dashed,black!45] (G)--(X) (G)--(Y) (G)--(Mp);
    \draw[fccblue,semithick] (X)--(M)--(Y)--(Xp)--(Mp)--(Yp)--cycle
      (Yp)--(Gp)--(M) (Gp)--(Xp);
    \draw[fccpair,densely dashed] (P)--(Q);
    \fill[black!65] (O) circle[radius=1.4pt];
    \node[anchor=north,fill=white,inner sep=1pt,font=\scriptsize]
      at (.50,.53,.43) {$\pi(1,1,1)$};
    \fill[fccpair] (P) circle[radius=2.3pt];
    \fill[fccpair] (Q) circle[radius=2.3pt];
    \node[fccpair,anchor=east,fill=white,inner sep=1.5pt]
      at (.70,-.04,.30) {$\bk$};
    \node[fccpair,anchor=west,fill=white,inner sep=1.5pt]
      at (.30,1.04,.70) {$\bG-\bk$};
    \foreach \v in {G,X,Y,M,Gp,Xp,Yp,Mp}
      \fill[fccblue] (\v) circle[radius=1.5pt];
    \node[anchor=north east,inner sep=2pt] at (G) {$\Gamma$};
    \node[anchor=north east,inner sep=3pt] at (X) {$X$};
    \node[anchor=north west,inner sep=3pt] at (Y) {$Y$};
    \node[anchor=north,inner sep=3pt] at (M) {$M$};
    \node[anchor=north west,inner sep=3pt] at (Gp) {$\Gamma'$};
    \node[anchor=south west,inner sep=3pt] at (Xp) {$X'$};
    \node[anchor=south east,inner sep=3pt] at (Yp) {$Y'$};
    \node[anchor=south east,inner sep=3pt] at (Mp) {$M'$};
  \end{scope}
  % Small coordinate triad, independent of the vertex labels.
  \begin{scope}[shift={(5.5,.52)}]
    \draw[->,black!65] (0,0)--(-.36,-.28) node[below,font=\scriptsize] {$k_x$};
    \draw[->,black!65] (0,0)--(.55,0) node[right,font=\scriptsize] {$k_y$};
    \draw[->,black!65] (0,0)--(0,.62) node[above,font=\scriptsize] {$k_z$};
  \end{scope}
  \node[align=center] at (3.1,-.60)
    {$\bG=2\pi(1,1,1)$\quad $\bk\sim\bG-\bk$ on the boundary};
\end{tikzpicture}
\caption{\label{fig:fcc:bz_rp2}
Fundamental domain of $F222$ with boundary identifications.
The domain $D$, chosen using reciprocal-lattice translations and $D_2$ rotations, occupies $1/4$ of the BZ volume.
In units of $2\pi$, the vertex coordinates are
$\Gamma=(0,0,0)$, $X=(1,0,0)$, $Y=(0,1,0)$, and $M=(1,1,0)$.
Primed vertices denote the antipodal partners of the corresponding unprimed vertices.
The three shaded faces are $F_i=\{\bk\in D:k_i=2\pi\}$ for $i=x,y,z$.
The union $F_x\cup F_y\cup F_z$ forms the surface on the $\Gamma'$ side of the closed path $X\to M\to Y\to X'\to M'\to Y'\to X$.
As illustrated by the two red points, the boundary identification pairs points antipodal about the center of the cube.}
\end{figure}

The key observation is that the irreducible Brillouin zone is equivalent to the three-dimensional real projective space $\R P^3$~\cite{ShiozakiAHSS2022}.
Figure~\ref{fig:fcc:bz_rp2} shows the fundamental domain
$D=[0,2\pi]^3$ under reciprocal-lattice translations and $D_2$ rotations.
Distinct points in the interior of $D$ are not identified, whereas boundary points are identified according to
\begin{align}
 \bk\sim\bG-\bk\qquad(\bk\in\partial D),\qquad
 \bG=2\pi(1,1,1).
 \label{eq:fcc:boundary_pairing}
\end{align}
For example, on the $k_z=0$ face,
$R_z(k_x,k_y,0)+\bG=\bG-(k_x,k_y,0)$.
Thus, a twofold rotation about the $z$ axis followed by a reciprocal-lattice translation maps the $k_z=0$ face to the $k_z=2\pi$ face centered on $\bG$.
The other two pairs of faces are related in the same way.
The gluing in Eq.~\eqref{eq:fcc:boundary_pairing} identifies points on $\p D$ that are antipodal about the cube center $\pi(1,1,1)$.

For the $2s$ orbitals at the atomic position $r$, we use the nonperiodic Bloch frame that includes the phase associated with the orbital position,
\begin{align}
 \Phi_r(\bk)=\sum_{\bm n\in L}e^{i\bk\cdot(\bm n+r)}|\bm n+r\rangle.
 \label{eq:fcc:frames}
\end{align}
The sewing matrices for the rotations $R_i$ are $I_2$, and $PT$ acts as $K$.
The frame obeys the quasiperiodicity condition
\begin{align}
 \Phi_r(\bk+2\pi\bm m)
 =e^{2\pi i\bm m\cdot r}\Phi_r(\bk),
 \qquad \bm m\in L^*.
 \label{eq:fcc:quasiperiodic_frame}
\end{align}

\noindent\textit{Proof of nonisomorphism in Proposition~\ref{prop:fcc:main}.} 
Suppose that a bundle isomorphism $E_{\mathcal B,\bk}\to E_{\mathcal A,\bk}$ exists,
and write the gauge transformation in the nonperiodic frames as $V(\bk)=\Phi_a(\bk)^\dagger\Phi_q(\bk)$.
PT symmetry and unitarity imply $V(\bk)\in O(2)$ and
\begin{align}
 V(R_i\bk)&=V(\bk),\qquad i=x,y,z,
 \label{eq:fcc:Q_rotation}\\
 V(\bk+2\pi\bm m)&=(-1)^{m_x}V(\bk),
 \qquad \bm m\in L^*.
 \label{eq:fcc:Q_period}
\end{align}
In particular, on the boundary of the BZ fundamental domain $D$,
\begin{align}
    V(\bG-\bk) = - V(\bk), \qquad \bk\in\partial D.
\end{align}
The $SO(2)$ rotation angle of $V(\bk)$ therefore has an odd winding number along the closed path $X \to M \to Y \to X' \to M' \to Y' \to X$ on $\p D$.
The rotation angle must then be discontinuous somewhere in the interior of $D$, so $V(\bk)$ cannot be defined over all of $D$.
\hfill$\square$\par

\subsection{Equivariant Euler class for two bands}
\label{subsec:fcc:twisted_euler}

We continue to consider spinless $Fm'm'm'$ with a trivial factor system.
Let $E$ be the rank-two real bundle fixed by $PT$.
We first construct its equivariant Euler class without assuming orientability.
We then restrict to the case in which all signs $\sigma_\gamma$, defined below, are $1$
and construct a gauge-invariant quantity that distinguishes $\mathcal A=2s_a$ from $\mathcal B=2s_b$.

\subsubsection{Integer cocycle from rotation angles}

We describe the BZ using the universal cover $\R^3$ and combine reciprocal-lattice translations and rotations into the group
\begin{align}
 \Gamma&:=L^*\rtimes D_2,\qquad
 (\bm m,r)\bk=r\bk+2\pi\bm m,\notag\\
 (\bm m,r)(\bm n,s)&=(\bm m+r\bm n,rs).
 \label{eq:fcc:euler_reciprocal_group}
\end{align}
The identity element of $\Gamma$ is $e=(\bm0,I)$.
The pullback of $E$ to $\R^3$ admits a global real orthonormal frame $u(\bk)$ because $\R^3$ is contractible.
A real frame on $\R^3$ need not be periodic, so no global real frame on the BZ is assumed.
The action $\widehat\gamma_{\bk}$ of $\gamma=(\bm m,r)$ is defined by the rotation $r$ together with the reciprocal-lattice identification.
We write the sewing matrix of $\widehat\gamma_{\bk}$ as
\begin{align}
 \widehat\gamma_{\bk}u(\bk)&=u(\gamma\bk)b_\gamma(\bk),\qquad
 b_\gamma(\bk)\in O(2),\notag\\
 b_{\gamma\delta}(\bk)&=b_\gamma(\delta\bk)b_\delta(\bk).
 \label{eq:fcc:twisted_euler_sewing}
\end{align}
Since $\R^3$ is connected, the sign $\sigma_\gamma:=\det b_\gamma(\bk)\in\{\pm1\}$ is independent of momentum.
The group multiplication law also gives $\sigma_{\gamma\delta}=\sigma_\gamma\sigma_\delta$.
Let $R(\theta)$ be the matrix for a two-dimensional rotation by $\theta$, and let $S:=\operatorname{diag}(1,-1)$ be the reflection matrix.
Because $\R^3$ is simply connected, we can choose continuous real-valued rotation angles such that
\begin{align}
 b_\gamma(\bk)&=R(\theta_\gamma(\bk))S^{\epsilon_\gamma},\qquad
 \sigma_\gamma=(-1)^{\epsilon_\gamma},\notag\\
 \epsilon_\gamma&\in\{0,1\},\qquad \theta_e(\bk)=0.
 \label{eq:fcc:euler_group_angle}
\end{align}
Using $SR(\theta)S=R(-\theta)$ and the multiplication law for the sewing matrices, we obtain the integer
\begin{align}
 n(\gamma,\delta)
 :=\frac{\theta_\gamma(\delta\bk)+\sigma_\gamma\theta_\delta(\bk)
          -\theta_{\gamma\delta}(\bk)}{2\pi}\in\Z.
 \label{eq:fcc:euler_group_cocycle}
\end{align}
The right-hand side is a continuous integer-valued function and is therefore independent of momentum.
The integer $n(\gamma,\delta)$ satisfies
\begin{align}
 \sigma_\gamma n(\delta,\zeta)-n(\gamma\delta,\zeta)
 +n(\gamma,\delta\zeta)-n(\gamma,\delta)&=0,\notag\\
 n(e,\gamma)=n(\gamma,e)&=0.
 \label{eq:fcc:euler_group_closed}
\end{align}
A change of angle branches, $\theta_\gamma\mapsto\theta_\gamma+2\pi m_\gamma$, gives
\begin{align}
 n(\gamma,\delta)&\longmapsto n(\gamma,\delta)+m_\gamma +\sigma_\gamma m_\delta-m_{\gamma\delta},\notag\\
 &m_\gamma\in\Z,\quad m_e=0.
 \label{eq:fcc:euler_group_branch}
\end{align}
We define the equivariant Euler class as the class of $n$ under the identification in Eq.~\eqref{eq:fcc:euler_group_branch},
\begin{align}
 e_{D_2}(E):=[n]\in H^2(\Gamma;\Z_\sigma).
 \label{eq:fcc:twisted_equivariant_euler}
\end{align}
Here, $\Z_\sigma$ is the coefficient system in which $\gamma$ acts on integers by multiplication by $\sigma_\gamma$.
Restriction to the translation subgroup $L^*$ gives the Euler class of $E$ over the BZ with coefficients in the orientation local system of $E$~\cite{MilnorCharacteristic1974}.

Under a frame change $u'=u h$ with $h(\bk)\in O(2)$,
\begin{align}
 b'_\gamma(\bk)=h(\gamma\bk)^{-1}b_\gamma(\bk)h(\bk).
 \label{eq:fcc:euler_group_gauge}
\end{align}
Connectedness of $\R^3$ implies that $\tau:=\det h(\bk)\in\{\pm1\}$ is constant, and $\sigma_\gamma$ is unchanged.
The terms arising from the rotation angle of $h(\bk)$ cancel in Eq.~\eqref{eq:fcc:euler_group_cocycle}, giving
\begin{align}
 n'(\gamma,\delta)=\tau n(\gamma,\delta)
                  +m_\gamma+\sigma_\gamma m_\delta-m_{\gamma\delta},\qquad m_e=0.
 \label{eq:fcc:euler_group_gauge_class}
\end{align}
Thus, $[n]$ and $-[n]$ cannot be distinguished unless an orientation of the frame is fixed.
The sign $\sigma$ and the unordered pair $\{[n],-[n]\}$ are independent of the frame.

\subsubsection{A gauge-invariant combination}

We now assume $\sigma_\gamma=1$ for every $\gamma\in\Gamma$, so that $E$ is orientable and the $D_2$ rotations preserve its orientation.
Under this assumption, we can construct an invariant from finitely many values of the cocycle $n$.
Define the group elements
\begin{align}
 t&:=((1,1,1),I),\qquad p:=(\bm0,R_x),\qquad q:=(\bm0,R_z),\notag\\
 g&:=tptq=((2,0,0),R_y),\qquad p^2=q^2=g^2=e,
 \label{eq:fcc:euler_relator_elements}
\end{align}
and consider the integers
\begin{align}
 S[n]&:=n(t,p)+n(tp,t)+n(tpt,q),\notag\\
 K[n]&:=2S[n]+n(g,g)-n(p,p)-n(q,q).
 \label{eq:fcc:euler_relator_integer}
\end{align}
Applying the transformation law in Eq.~\eqref{eq:fcc:euler_group_gauge_class}, we find
\begin{align}
 S[n']&=\tau S[n]+2m_t+m_p+m_q-m_g,\notag\\
 K[n']&=\tau K[n]+4m_t.
 \label{eq:fcc:euler_relator_gauge}
\end{align}
Therefore,
\begin{align}
 I(E):=K[n]\bmod4\in\Z_4
 \label{eq:fcc:euler_relator_mod4}
\end{align}
is independent of the choice of angle branches and changes sign when the frame orientation is reversed.
If no orientation is specified, the invariant is the unordered pair $\{I(E),-I(E)\}$,
which distinguishes the three possibilities $0$, $\{1,3\}$, and $2$.
The invariant $\{I(E),-I(E)\}$ is defined for any real two-band bundle satisfying $\sigma_\gamma=1$.

\subsubsection{Comparison of the two configurations}

In the real frames of Eq.~\eqref{eq:fcc:frames}, which include the position-dependent phases,
\begin{align}
 b^{2s_a}_{(\bm m,r)}=I_2,\qquad
 b^{2s_b}_{(\bm m,r)}=(-1)^{m_x}I_2.
 \label{eq:fcc:euler_atomic_sewing}
\end{align}
The configurations $2s_a$ and $2s_b$ both have $\sigma_\gamma=1$, so the invariant $\{I(E),-I(E)\}$ is defined.
For $2s_a$, all angles can be chosen to vanish, giving $n_{2s_a}=0$.
For $2s_b$, choosing $\theta_{(\bm m,r)}=\pi m_x$ gives
\begin{align}
 n_{2s_b}\bigl((\bm m,r),(\bm v,s)\bigr)
 =\frac{v_x-(r\bm v)_x}{2}.
 \label{eq:fcc:euler_atomic_mixed}
\end{align}
Substitution into the above combinations yields
\begin{align}
 S[n_{2s_b}]&=0,\qquad n_{2s_b}(g,g)=2,\notag\\
 n_{2s_b}(p,p)&=n_{2s_b}(q,q)=0,
 \label{eq:fcc:euler_atomic_evaluation}
\end{align}
and hence
\begin{align}
 I(2s_a)=0,\qquad I(2s_b)=2.
 \label{eq:fcc:euler_atomic_obstruction}
\end{align}
The occupied bundles of the two configurations are therefore not equivariantly isomorphic.

Moreover, the integer 1-cochain $c(\bm m,r):=m_x$ satisfies
\begin{align}
 2n_{2s_b}(\gamma,\delta)=c(\gamma)+c(\delta)-c(\gamma\delta).
 \label{eq:fcc:euler_atomic_order_two}
\end{align}
The nonzero class $[n_{2s_b}]$ thus has order two.
The ordinary Euler classes of both configurations vanish because Eq.~\eqref{eq:fcc:euler_atomic_mixed} vanishes when restricted to translations.

\section{Spinless \texorpdfstring{$P\overline1'$}{P-1'}: adiabatic deformations beyond stable Wyckoff equivalence}
\label{sec:WCP_move}

We construct two four-orbital configurations $\mathcal A$ and $\mathcal B$ in spinless $P\overline1'$ that are adiabatically equivalent but not stably Wyckoff equivalent.
We first give an adiabatic deformation in a fixed eight-orbital model.
We then relate the nonuniqueness of the Wannier basis to a Chern-insulator pump and discuss the closing of the surface gap.

\subsection{Two atomic configurations}

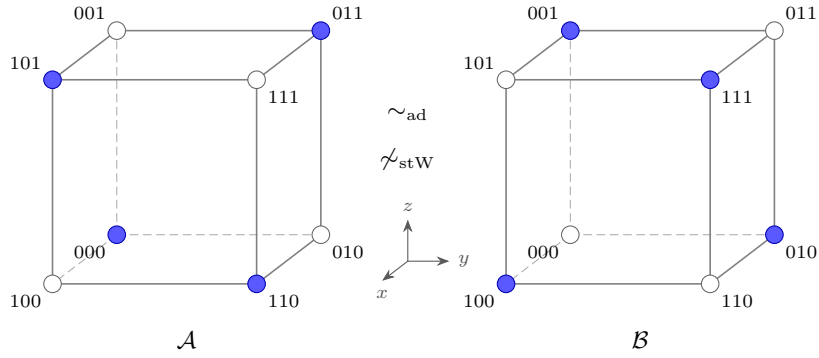
\begin{figure*}[t]
    \centering
    \begin{tikzpicture}[font=\small,>=Stealth,line join=round,line cap=round]
        \foreach \offset/\ptparity/\ptname in {0/0/{\mathcal A},6/1/{\mathcal B}}{
            \begin{scope}[shift={(\offset cm,0cm)}]
                \begin{scope}[x={(-.85cm,-.65cm)},y={(2.7cm,0cm)},z={(0cm,2.7cm)}]
                    % Coordinates are twice the real-space position.
                    \draw[densely dashed,gray!65]
                        (0,0,0)--(1,0,0) (0,0,0)--(0,1,0) (0,0,0)--(0,0,1);
                    \draw[gray,semithick]
                        (1,0,0)--(1,1,0)--(0,1,0)--(0,1,1)--(0,0,1)--(1,0,1)--cycle
                        (1,0,1)--(1,1,1)--(1,1,0) (1,1,1)--(0,1,1);
                    \foreach \xx in {0,1}{\foreach \yy in {0,1}{\foreach \zz in {0,1}{
                        \pgfmathtruncatemacro{\ptvertexparity}{mod(\xx+\yy+\zz,2)}
                        \ifnum\ptvertexparity=\ptparity
                            \filldraw[fill=blue!65,draw=blue!70!black]
                                (\xx,\yy,\zz) circle[radius=3.3pt];
                        \else
                            \filldraw[fill=white,draw=gray!80!black]
                                (\xx,\yy,\zz) circle[radius=3.3pt];
                        \fi
                    }}}
                    \node[anchor=north east,inner sep=4pt,font=\scriptsize] at (0,0,0) {$000$};
                    \node[anchor=north east,inner sep=4pt,font=\scriptsize] at (1,0,0) {$100$};
                    \node[anchor=north west,inner sep=4pt,font=\scriptsize] at (0,1,0) {$010$};
                    \node[anchor=north west,inner sep=4pt,font=\scriptsize] at (1,1,0) {$110$};
                    \node[anchor=south east,inner sep=4pt,font=\scriptsize] at (0,0,1) {$001$};
                    \node[anchor=south east,inner sep=4pt,font=\scriptsize] at (1,0,1) {$101$};
                    \node[anchor=south west,inner sep=4pt,font=\scriptsize] at (0,1,1) {$011$};
                    \node[anchor=north west,inner sep=4pt,font=\scriptsize] at (1,1,1) {$111$};
                    \node[below=10pt] at (.5,.5,-.18) {$\ptname$};
                \end{scope}
            \end{scope}
        }
        \node at (3.85,1.6) {$\sim_{\mathrm{ad}}$};
        \node at (3.85,1.0) {$\not\sim_{\mathrm{stW}}$};
        \begin{scope}[shift={(3.85,-.35)}]
            \draw[->,black!65] (0,0)--(-.34,-.26) node[below,font=\scriptsize] {$x$};
            \draw[->,black!65] (0,0)--(.55,0) node[right,font=\scriptsize] {$y$};
            \draw[->,black!65] (0,0)--(0,.55) node[above,font=\scriptsize] {$z$};
        \end{scope}
    \end{tikzpicture}
    \caption{Two atomic configurations in spinless $P\overline1'$.
    Each cube represents $[0,\frac12]^3$, and the vertex label $\epsilon_x\epsilon_y\epsilon_z$ denotes the position $(\epsilon_x,\epsilon_y,\epsilon_z)/2$.
    There is one orbital at each vertex. Filled blue circles denote occupied orbitals, and open circles denote the unoccupied orbitals used in the adiabatic deformation.
    The four orbitals at even-parity vertices are occupied in $\mathcal A$, whereas the four orbitals at odd-parity vertices are occupied in $\mathcal B$.
    The configurations $\mathcal A$ and $\mathcal B$ are not stably Wyckoff equivalent, but they are connected by an adiabatic deformation within the eight-orbital space $\mathcal A\oplus\mathcal B$, with four bands occupied throughout.}
    \label{fig:pt:configurations}
\end{figure*}

Let the lattice be $\Z^3$, with PT acting as $\bx\mapsto-\bx$ in real space.
Modulo translations, there are eight PT-fixed points $\bw\in\{0,\frac12\}^{\times3}$.
Since $(\widehat{\mathrm{PT}})^2=1$, each fixed point has a single one-dimensional irreducible corepresentation.
We denote the corresponding occupied orbital by $s_{\bw}$.
Consider the two configurations in Fig.~\ref{fig:pt:configurations},
\begin{align}
    \mathcal A
    &=s_{000}\oplus s_{\frac12\frac12 0}
      \oplus s_{\frac12 0\frac12}\oplus s_{0\frac12\frac12},\nonumber\\
    \mathcal B
    &=s_{\frac12 00}\oplus s_{0\frac12 0}
      \oplus s_{00\frac12}\oplus s_{\frac12\frac12\frac12}.
\end{align}
A PT-related pair at general positions can be moved to any fixed point, giving the Wyckoff equivalences
\begin{align}
    s_{\bw_{\rm gp}}\oplus s_{-\bw_{\rm gp}}
    \sim_{\mathrm W}2s_{\bw}
    \sim_{\mathrm W}2s_{\bw'}.
    \label{eq:WCP_move_2}
\end{align}
Moving orbitals away from a fixed point requires a PT-related pair, so a Wyckoff deformation preserves the parity of the number of occupied orbitals at each fixed point.
The two configurations have different parities at all eight points. These differences persist after adding a common occupied atomic ancilla, so
\begin{align}
    \mathcal A\not\sim_{\mathrm{stW}}\mathcal B.
\end{align}
We now show that
\begin{align}
    \mathcal A\sim_{\mathrm{ad}}\mathcal B.
    \label{eq:pt_four_relations}
\end{align}

\subsection{Adiabatic deformation with eight fixed orbitals}
\label{sec:adia_2}

We use the eight orbitals of $\mathcal A\oplus\mathcal B$ as a basis for the full orbital space.
In the periodic Bloch basis $\ket{\bk,\bw}=\sum_{\bR}e^{i\bk\cdot\bR}\ket{\bR+\bw}$,
PT acts on $s_{\bw}$ as $e^{2i\bw\cdot\bk}K$, where $K$ denotes complex conjugation.
With the orbital ordering above, the PT matrices of the two configurations are
\begin{align}
    u^{\mathcal A}_{\bk}
    &=
    \operatorname{diag}
    \bigl(
        1,
        e^{i(k_x+k_y)},
        e^{i(k_x+k_z)},
        e^{i(k_y+k_z)}
    \bigr),
    \nonumber\\
    u^{\mathcal B}_{\bk}
    &=
    \operatorname{diag}
    \bigl(
        e^{ik_x},
        e^{ik_y},
        e^{ik_z},
        e^{i(k_x+k_y+k_z)}
    \bigr).
    \label{eq:U1U2_def}
\end{align}
On the full space, $U_\bk=\operatorname{diag}(u^{\mathcal A}_{\bk},u^{\mathcal B}_{\bk})$,
and PT symmetry requires $U_\bk H_t(\bk)^*U_\bk^\dagger=H_t(\bk)$.

Suppose that a periodic matrix $V_\bk$, invertible throughout the BZ, satisfies
\begin{align}
    u^{\mathcal A}_{\bk}V_\bk^*\bigl(u^{\mathcal B}_{\bk}\bigr)^\dagger=V_\bk.
    \label{eq:PT_for_V}
\end{align}
Using the construction in Proposition~\ref{prop:atomic_ad_intertwiner}, we define the Hamiltonian
\begin{align}
    H_t(\bk)
    &=\cos(\pi t)
    \begin{pmatrix}
        -I_4&0\\
        0&I_4
    \end{pmatrix}\notag\\
    &\quad+\sin(\pi t)
    \begin{pmatrix}
        0&V_\bk\\
        V_\bk^\dagger&0
    \end{pmatrix},\qquad 0\leq t\leq1.
    \label{eq:adia_model}
\end{align}
The occupied configuration is $\mathcal A$ at $t=0$ and $\mathcal B$ at $t=1$.
If $s_j(\bk)>0$ are the singular values of $V_\bk$, the energy eigenvalues are
\begin{align}
    E_{j,\pm}(\bk,t)
    =
    \pm
    \sqrt{
        \cos^2(\pi t)
        +
        s_j(\bk)^2\sin^2(\pi t)
    },
\end{align}
so the bulk gap remains open along the entire path.

\begin{figure*}[t]
    \centering
    \resizebox{\linewidth}{!}{%
    \begin{tikzpicture}[x=1bp,y=-1bp]
        \path[use as bounding box] (0,0) rectangle (850.32,226.8);
        % Move the original right panel to the left, retaining the TeX orbital labels.
        \begin{scope}[overlay,xshift=-605bp]
            \clip (605,0) rectangle (850.32,226.8);
        \node[anchor=north west,inner sep=0pt] at (0,0)
            {\includegraphics{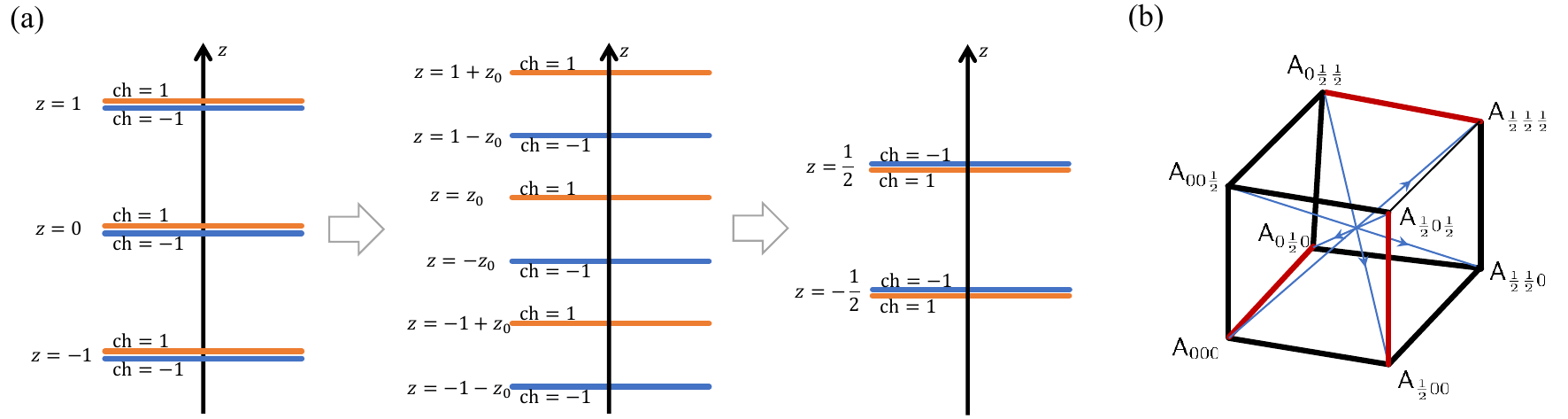}};
        % The imported PDF has outlined orbital labels; replace them in TeX.
        \foreach \px/\py/\pw/\psub in {
            697/29/34/{0\frac12\frac12},
            808/54/34/{\frac12\frac12\frac12},
            632/86/32/{00\frac12},
            682/120/30/{0\frac12 0},
            758/112/34/{\frac12 0\frac12},
            808/141/34/{\frac12\frac12 0},
            635/182/28/{000},
            757/200/31/{\frac12 00}}{
            \fill[white] (\px,\py) rectangle ++(\pw,20);
            \node[anchor=base west,inner sep=0pt,
                  font=\fontsize{14}{14}\selectfont]
                at (\px,{\py+11}) {$s_{\psub}$};
        }
            \fill[white] (608,0) rectangle (634,21);
            \node[anchor=north west,inner sep=0pt,
                  font=\fontsize{16.8}{18}\selectfont] at (611.9,1.2) {(a)};
        \end{scope}
        % Move the Chern-pump panel to the right.
        \begin{scope}[overlay,xshift=245.32bp]
            \clip (0,0) rectangle (605,226.8);
            \node[anchor=north west,inner sep=0pt] at (0,0)
                {\includegraphics{adia_1.pdf}};
            \fill[white] (2,0) rectangle (28,21);
            \node[anchor=north west,inner sep=0pt,
                  font=\fontsize{16.8}{18}\selectfont] at (5.45,1.2) {(b)};
        \end{scope}
    \end{tikzpicture}%
    }

    \caption{Two PT-symmetric adiabatic deformations.
    (a) Hopping amplitudes in the model connecting $\mathcal A$ and $\mathcal B$ in a fixed eight-orbital space.
    The cube represents $[0,1/2]^3$. Black and red lines denote coefficients $+1$ and $-1$ in $V_{0,\bk}$, respectively.
    Blue arrows denote the hoppings proportional to $i\lambda$ added to $V_\bk$.
    (b) Chern insulator pump. Orange and blue lines denote layers with Chern numbers $+1$ and $-1$, respectively.
    For $0\leq t\leq1/2$, the layers move from $z=n$ ($n\in\Z$) to $z=n+t$ and $z=n-t$.
    At $t=1/2$, layers with opposite Chern numbers coincide again at half-integer values of $z$.}

    \label{fig:adia}
\end{figure*}

We now construct $V_\bk$ explicitly.
The corresponding real-space hoppings are shown in Fig.~\ref{fig:adia}(a).
As in the octupole-insulator model~\cite{BenalcazarElectric2017}, we first choose hoppings that give a $\pi$ flux through each face of the cube and set
\begingroup
\setlength{\arraycolsep}{2.5pt}
\begin{align}
    V_{0,\bk}
    =
    \begin{pmatrix}
        1+e^{-ik_x}&-1-e^{-ik_y}&1+e^{-ik_z}&0\\
        1+e^{ik_y}&1+e^{ik_x}&0&1+e^{-ik_z}\\
        -1-e^{ik_z}&0&1+e^{ik_x}&1+e^{-ik_y}\\
        0&1+e^{ik_z}&1+e^{ik_y}&-1-e^{-ik_x}
    \end{pmatrix}.
\end{align}
\endgroup
The determinant of $V_{0,\bk}$ is
\begin{align}
    \det V_{0,\bk}
    =
    -4(3+\cos k_x+\cos k_y+\cos k_z)^2,
\end{align}
so the gap closes only at $\bk=(\pi,\pi,\pi)$.
To remove the zero at $\bk=(\pi,\pi,\pi)$, we take
\begin{widetext}
\begin{align}
    V_\bk
    =
    V_{0,\bk}
    +
    \lambda
    \begin{pmatrix}
        0&0&0&i-i e^{-i(k_x+k_y+k_z)}\\
        0&0&-i+i e^{i(k_x+k_y-k_z)}&0\\
        0&i-i e^{i(k_x-k_y+k_z)}&0&0\\
        i-i e^{i(-k_x+k_y+k_z)}&0&0&0
    \end{pmatrix},
    \label{eq:LGT_ex}
\end{align}
\end{widetext}
where $\lambda\in\R$.
% \footnote{\unexpanded{The expansion to first order near the degeneracy is
% \begin{align*}
% V_{0,(\pi,\pi,\pi)+\delta\bk}
% &=\sum_\mu\delta k_\mu M_\mu\notag\\
% &\quad+O(|\delta\bk|^2),
% \end{align*}
% with
% \begin{align*}
% M_x&=\begin{pmatrix}
%     i\\
%     &-i\\
%     &&-i\\
%     &&&-i\\
% \end{pmatrix},\\
% M_y&=\begin{pmatrix}
%     &-i\\
%     -i\\
%     &&&i\\
%     &&-i\\
% \end{pmatrix},\\
% M_z&=\begin{pmatrix}
%     &&i\\
%     &&&i\\
%     i\\
%     &-i\\
% \end{pmatrix}.
% \end{align*}
% To add a mass $X$ to the off-diagonal block of the corresponding Hermitian matrix, it is enough to choose an invertible matrix $X$ satisfying
% \begin{align*}
%     M_\mu X^\dagger+X M_\mu^\dagger&=0,\\
%     M_\mu^\dagger X+X^\dagger M_\mu&=0,\qquad \mu=x,y,z.
% \end{align*}
% For example,
% \begin{align*}
%     X=\begin{pmatrix}
%         &&&i\\
%         &&-i\\
%         &i\\
%         i\\
%     \end{pmatrix}
% \end{align*}
% satisfies the mass conditions and $XX^\dagger=I_4$.
% Symmetrizing the mass term $X$ under PT gives the second term in Eq.~\eqref{eq:LGT_ex}.
% }}
The matrix $V_\bk$ in Eq.~\eqref{eq:LGT_ex} satisfies Eq.~\eqref{eq:PT_for_V} and is invertible throughout the BZ for $0<|\lambda|<\sqrt2$.%
% \footnote{\unexpanded{Indeed, setting $t_\mu=\cos k_\mu$, we can write the determinant as
% \begin{align*}
%     \det V_\bk
%     =
%     -4\Bigl\{
%         &
%         \bigl[
%             (1-\lambda^2)(3+t_x+t_y+t_z)
%             +2\lambda^2
%         \bigr]^2
%         \nonumber\\
%         &+
%         2\lambda^2(3-\lambda^2)
%         (1+t_x)(1+t_y)(1+t_z)
%     \Bigr\}.
% \end{align*}
% For $0<\lambda^2<2$, the second term in braces is nonnegative.
% If the second term vanishes, at least one $t_\mu$ equals $-1$, so $0\leq3+t_x+t_y+t_z\leq4$.
% For $0\leq3+t_x+t_y+t_z\leq4$, the expression squared in the first term is at least $2\lambda^2$ for $0<\lambda^2\leq1$ and at least $4-2\lambda^2$ for $1<\lambda^2<2$.
% Both lower bounds are positive, and hence $\det V_\bk<0$.
% }}

Thus, Eq.~\eqref{eq:adia_model} gives a PT-symmetric adiabatic deformation from $\mathcal A$ to $\mathcal B$ with four occupied bands throughout.

\subsection{Wannier bases and large gauge transformations}
\label{sec:adia_wannier_center}

\begin{figure}[!t]
    \centering
    \includegraphics[width=0.60\linewidth]{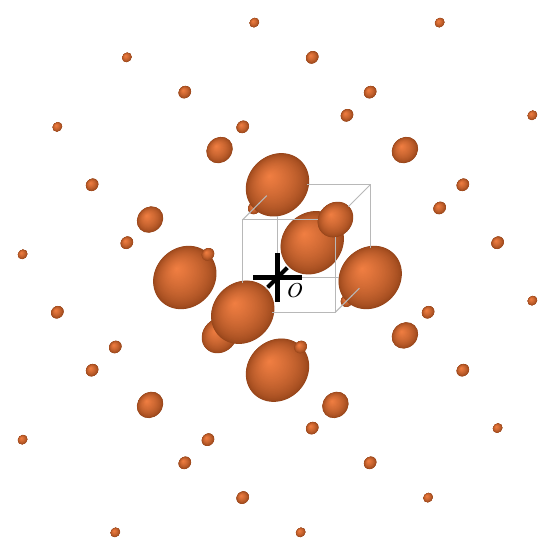}
    \caption{Probability distribution of the Wannier state $\ket{W_{1\bm0}(1)}$ at $t=1$ in the $\Phi_t(\bk)$ gauge ($n=1$, $\lambda=1$).
    The volume of each orange sphere is proportional to the probability on the corresponding orbital of $\mathcal B$.
    The black cross marks the Wannier center, which remains at the origin.
    The gray edges enclose $[0,1/2]^3$, one eighth of a unit cell, and $O$ denotes the origin.}
    \label{fig:pt_wannier}
\end{figure}

In Eq.~\eqref{eq:adia_model}, replace $V_\bk$ by its unitary polar factor,
\begin{align}
    \bar V_\bk=V_\bk(V_\bk^\dagger V_\bk)^{-1/2}.
\end{align}
The replacement $V_\bk\mapsto\bar V_\bk$ leaves the endpoints unchanged and preserves the symmetry and the bulk gap.
An occupied frame for the resulting model is
\begin{align}
    \Phi_t(\bk)
    &=\begin{pmatrix}
        \cos\frac{\pi t}{2}\,I_4\\
        -\sin\frac{\pi t}{2}\,\bar V_\bk^\dagger
      \end{pmatrix}, &
    U_\bk\Phi_t(\bk)^*&=\Phi_t(\bk)u^{\mathcal A}_{\bk}.
\end{align}
Since $\bar V_\bk$ is periodic and analytic, the Fourier transform in Eq.~\eqref{eq:atomic_ad_wannier_components} is exponentially localized.
The sewing matrix with diagonal entries $e^{2i\bw_n\cdot\bk}$ gives the transformation law
\begin{align}
    \widehat{\mathrm{PT}}\ket{W_{n\bR}}
    =\ket{W_{n,-\bR-2\bw_n}}.
    \label{eq:PT_Wannier}
\end{align}
The Wannier centers are therefore fixed at $\bR+\bw_n$.
Thus, the Wannier centers of $\Phi_t$ remain at the positions of $\mathcal A$ along the entire path.
At $t=1$, the Wannier states have support only on the orbitals of $\mathcal B$. Each state remains spread over several orbitals, with its center at a position in $\mathcal A$ (Fig.~\ref{fig:pt_wannier}).

Another frame for the same occupied subspace is
\begin{align}
    \Psi_t(\bk)
    &=\Phi_t(\bk)\bar V_\bk
      =\begin{pmatrix}
        \cos\frac{\pi t}{2}\,\bar V_\bk\\
        -\sin\frac{\pi t}{2}\,I_4
      \end{pmatrix},\notag\\
    U_\bk\Psi_t(\bk)^*&=\Psi_t(\bk)u^{\mathcal B}_{\bk}.
\end{align}
The Wannier centers of $\Psi_t(\bk)$ remain at the positions of $\mathcal B$ throughout the path.
Hence, the same occupied subspace admits two Wannier bases whose atomic configurations are not stably Wyckoff equivalent.

The three-dimensional winding number of the gauge transformation relating the two frames is
\begin{align}
    W_3[V]
    :=\frac{1}{24\pi^2}\int_{T^3}\tr\bigl[(V_\bk^{-1}dV_\bk)^3\bigr]
    =W_3[\bar V]\in\Z.
\end{align}
For the matrix in Eq.~\eqref{eq:LGT_ex}, direct calculation gives
\begin{align}
    W_3[V]=\sgn(\lambda),\qquad 0<|\lambda|<\sqrt2.
\end{align}
Moreover, every invertible matrix satisfying Eq.~\eqref{eq:PT_for_V} has an odd winding number~\cite{Shiozaki$mathbbZ_2$2025}.
Thus, the large gauge transformation $\bar V_\bk$ between the two configurations cannot be contracted to the identity.

\subsection{Interpretation as a Chern insulator pump}
\label{sec:adia_1}

Explicit cyclic pumps have been described using hybrid Wannier sheets and coupled Chern-insulator layers~\cite{TaherinejadAdiabatic2015,OlsenSurface2017}.
The equivalence between $\mathcal A$ and $\mathcal B$ can also be understood in terms of the Chern insulator pump in Fig.~\ref{fig:adia}(b).
Place four orbitals at the origin of each unit cell and consider the following model in the $z=0$ plane:
\begin{align}
    H(\bk)
    &=\sin k_x\,\sigma_x-\sin k_y\,\sigma_y\tau_y\notag\\
    &\quad+(1-\cos k_x-\cos k_y)\sigma_z,\qquad U_\bk=I_4,
    \label{eq:P-1'_FI_model}
\end{align}
where $\sigma_\mu$ and $\tau_\mu$ are two sets of Pauli matrices.
The matrix $H(\bk)$ is real.
The $\tau_y=\pm1$ subspaces form layers with opposite Chern numbers $\pm1$ and are exchanged by PT.
Denoting the occupied states by $\mathrm{BI}(H)$, we have the $K$-group relation
\begin{align}
    [\mathrm{BI}(H)]+[s_{000}]
    =[s_{\frac12\frac12 0}]+[s_{\frac12 00}]+[s_{0\frac12 0}],
    \label{eq:z0_equ_rel}
\end{align}
which expresses a relation associated with fragile topology in terms of atomic configurations~\cite{PoFragile2018,AhnFailure2019,WadaGeneral2024}.
Square brackets denote elements of the $K$-group.

Move the two layers to $z=\pm t$ and vary $t$ from $0$ to $\frac12$.
The two Chern insulator layers then meet at $z=\frac12$, modulo translations.
The intralayer hoppings remain unchanged, so Eq.~\eqref{eq:z0_equ_rel} and the corresponding relation for the layers translated to $z=\frac12$ give
\begin{align}
    &[s_{\frac12\frac12 0}]+[s_{\frac12 00}]+[s_{0\frac12 0}]-[s_{000}]
    \nonumber\\
    &\qquad=[s_{\frac12\frac12\frac12}]+[s_{\frac12 0\frac12}]
              +[s_{0\frac12\frac12}]-[s_{00\frac12}].
\end{align}
Combining the layer-pump relation with Eq.~\eqref{eq:WCP_move_2} gives $\mathcal A\sim_{\mathrm{st\,ad}}\mathcal B$.
The Chern insulator pump corresponds to the third differential of the real-space AHSS~\cite{ShiozakiGeneralized2023}.
The $K$-group relation establishes stable adiabatic equivalence, whereas the eight-orbital model in Sec.~\ref{sec:adia_2} connects the same two configurations without adding occupied ancillas.

\subsection{The spin-CS term and surface gap closing}
\label{sec:sCS}

Although the two configurations can be connected without closing the bulk gap, the surface gap must close along the path when the boundary termination is fixed.
To establish this result, we use the $4\pi$-periodic spin Chern--Simons (spin-CS) term defined by a connection on the PT-real bundle~\cite{JenquinClassical2005,Shiozaki$mathbbZ_2$2025}.
The spin-CS term is a gauge-invariant geometric quantity that depends on the connection and is not quantized by PT symmetry alone.

On each patch, choose a local frame satisfying $U_\bk(\Phi^\alpha)^*=\Phi^\alpha$.
Starting from the ordinary Berry connection $(A_\Phi)^\alpha=(\Phi^\alpha)^\dagger d\Phi^\alpha$, define the averaged real connection~\cite{DeNittisDifferential2016} by
\begin{align}
    (B_\Phi)^\alpha
    &:=(\Phi^\alpha)^\dagger
      \left(d+\frac12U_\bk dU_\bk^\dagger\right)\Phi^\alpha
      \nonumber\\
    &=\frac12\left[(A_\Phi)^\alpha+\bigl((A_\Phi)^\alpha\bigr)^*\right].
    \label{eq:PT_BC}
\end{align}
The averaged connection $B_\Phi$ is a real antisymmetric matrix-valued one-form and defines an $O(4)$ connection on the occupied real bundle.
For a metric $g$ and a spin structure $\sigma$ on the BZ, define the spin-CS term using the reduced $\eta$ invariant of the Dirac operator coupled to $B_\Phi$~\cite{AtiyahSpectral1975,JenquinClassical2005}:
%\footnote{\unexpanded{The function $\eta_{iD}(s)=\sum_{\lambda\ne0}\sgn(\lambda)|\lambda|^{-s}$ is analytically continued to $s=0$ from sufficiently large $\operatorname{Re}s$.
% Eigenvalues $\lambda$ are counted with multiplicity.
% The three-dimensional Dirac operator coupled to a real bundle has a twofold degeneracy, so $\bar\eta$ varies smoothly modulo $2$.
% The quantity $\bar\eta_{\rm triv}$ refers to the trivial real line bundle.
% Subtracting the rank times $\bar\eta_{\rm triv}$ removes the metric dependence by the APS index theorem.}}
\begin{align}
    \bar\eta(B_\Phi,g,\sigma)
    &:=\frac{\eta_{iD_{B_\Phi}}(0)+\dim\ker(iD_{B_\Phi})}{2}
      \quad\bmod2,\\
    \sCS(\Phi,\sigma)
    &:=2\pi\bigl[\bar\eta(B_\Phi,g,\sigma)
        -4\bar\eta_{\rm triv}(g,\sigma)\bigr]
      \quad\bmod4\pi.
\end{align}
The spin structure here defines the Dirac operator on the BZ and is unrelated to the electron spin.

Let $\sigma_{\mathbb{NS}}$ be the spin structure with antiperiodic boundary conditions in all three directions.
At the atomic limits~\cite{Shiozaki$mathbbZ_2$2025},
%\footnote{\unexpanded{For a single orbital $s_{\bw}$, the local connection in the PT-real frame $e^{i\bw\cdot\bk}$ vanishes.
% The signs acquired on going around the BZ shift the spin structure as $\sigma\mapsto\sigma+2\bw$.
% Thus,
% \begin{align*}
%     \sCS(s_{\bw},\sigma)
%     \equiv2\pi\bigl[\bar\eta_{\rm triv}(g,\sigma+2\bw)
%                      -\bar\eta_{\rm triv}(g,\sigma)\bigr]\pmod{4\pi}.
% \end{align*}
% On the flat torus $T^3$, the nonzero eigenvalues cancel in pairs of opposite sign.
% Thus, $\bar\eta_{\rm triv}=1$ only for periodic boundary conditions in all directions and is zero otherwise.
% For $\sigma_{\mathbb{NS}}$, only $\bw=(\frac12,\frac12,\frac12)$ contributes $2\pi$.
% Adding the contributions of the independent atomic orbitals gives the stated result.}}
\begin{align}
    \sCS(\mathcal A,\sigma_{\mathbb{NS}})&\equiv0, &
    \sCS(\mathcal B,\sigma_{\mathbb{NS}})&\equiv2\pi\pmod{4\pi}.
\end{align}

The different spin-CS values reflect different connections and are consistent with the PT-symmetric adiabatic deformation constructed in Sec.~\ref{sec:adia_2}.

\begin{prop}
Fix the PT action $U_\bk$ on the eight-orbital space $\mathcal A\oplus\mathcal B$.
For any smooth PT-symmetric adiabatic deformation connecting the atomic limits $\mathcal A$ and $\mathcal B$ with four occupied bands, the relative second Chern number defined using the atomic-orbital bases as trivializations at the two endpoints is odd.
\end{prop}

\noindent\textit{Proof.} 
Let $F$ be the ordinary Berry curvature.
Since the Berry connection vanishes in the atomic-orbital bases at both endpoints, the relative second Chern number is
\begin{align}
    \operatorname{ch}_2
    =-\frac{1}{8\pi^2}\int_{T^3\times[0,1]}\tr(F\wedge F)\in\Z.
\end{align}
Using the real connection $B_\Phi$ gives the same integral.%
\footnote{\unexpanded{The difference between the two Chern--Weil forms is exact.
In the atomic-orbital bases at the endpoints, the ordinary connection is $A=0$, while the real connection is represented by $\tilde B=-i\operatorname{diag}(\bw_n\cdot d\bk)$.
The interpolating connections $s\tilde B$, $s\in[0,1]$, are flat, so the transgression term vanishes.}}
The variation formula from the APS index theorem~\cite{JenquinClassical2005} gives
\begin{align}
    \operatorname{ch}_2
    \equiv\frac{\sCS(\mathcal B,\sigma_{\mathbb{NS}})
                 -\sCS(\mathcal A,\sigma_{\mathbb{NS}})}{2\pi}
    \equiv1\pmod2.
\end{align}
The $4\pi$ periodicity of the spin-CS term fixes the parity of the relative second Chern number.
\hfill$\square$\par

By bulk--boundary correspondence~\cite{TaherinejadAdiabatic2015,OlsenSurface2017}, the surface gap must close along any adiabatic path connecting the atomic configurations ${\cal A}$ and ${\cal B}$.
Weyl points appear in the space spanned by the two surface momenta $\bk_{\parallel}$ and $t$, with an odd total Weyl charge.
The need for a gap closing at a fixed boundary despite a gapped bulk interpolation is known as a boundary obstruction~\cite{KhalafBoundary2021}.

A smooth periodic occupied frame can be chosen along the entire adiabatic path.
However, the nonzero relative second Chern number prevents a continuous frame from matching the specified atomic-limit frames at both endpoints.
For example, $\Phi_t$ matches the atomic-limit frame of $\mathcal A$ at $t=0$, but at $t=1$ it differs from the atomic-limit frame of $\mathcal B$ by a gauge transformation with nonzero three-dimensional winding number.

The situation changes if we also impose a symmetry that reverses spacetime orientation, such as time reversal, inversion, or reflection.
For a spin structure $\sigma$ preserved by the additional spacetime-orientation-reversing symmetry,
\begin{align}
    \sCS(B,\sigma)\equiv-\sCS(B,\sigma)\pmod{4\pi},
\end{align}
so the value is quantized to $0$ or $2\pi$.
If $\mathcal A$ and $\mathcal B$ both obey this additional symmetry but have different spin-CS values, any adiabatic deformation connecting them must break it.
Indeed, the term proportional to $\lambda$ in Eq.~\eqref{eq:LGT_ex} violates the time-reversal condition $V_{-\bk}=V_\bk^*$.

\section{Classification of stable equivalence}
\label{sec:classification}
\label{clsnew:classification}

Section~\ref{sec:WCP_move} showed that an adiabatic deformation can connect two configurations that are not stably Wyckoff equivalent.
Here we classify when stable Wyckoff equivalence and stable adiabatic equivalence coincide.
We consider charge-conserving, noninteracting electron systems with three-dimensional magnetic space groups.
We use the trivial factor system for spinless electrons and the electronic spin-$1/2$ factor system for spinful electrons.
Extensions to general factor systems and spin space groups are beyond the scope of this section.

In the real-space AHSS, the difference between the two stable equivalence relations is described by the image of the third differential.
We first explain how the third differential relates the two classifications, then determine it using symmetric models and the known $E^2$ data.

For an element $g=\{p_g|\bt_g\}$ of the magnetic space group, set $\phi_g=+1$ if $g$ is unitary and $\phi_g=-1$ if $g$ is antiunitary.
For a matrix $M$, we use the notation
\begin{align}
    M^{\phi_g} = \begin{cases}
        M,&\phi_g=+1,\\
        M^*,&\phi_g=-1
    \end{cases}.
\end{align}
The internal action is written as $u(g)K^{(1-\phi_g)/2}$, where $u(g)$ denotes the matrix part of the symmetry action and $K$ denotes complex conjugation.

\subsection{Third differential}
\label{clsnew:first}
\label{clsnew:higher}

The group $E^2_{0,0}$ of the real-space AHSS consists of formal differences of atomic configurations,
with the difference of any two configurations connected by a Wyckoff deformation set to zero.
The group $E^2_{0,0}$ is the group completion of the commutative monoid of Wyckoff equivalence classes under direct sum.
Writing the image of an atomic configuration $\mathcal A$ in $E^2_{0,0}$ as $[\mathcal A]$, the definition of group completion gives
\begin{align}
 [\mathcal A]=[\mathcal B]\in E^2_{0,0}
 \quad\Longleftrightarrow\quad
 \mathcal A\sim_{\mathrm{stW}}\mathcal B.
 \label{clsnew:E2-stW}
\end{align}
Comprehensive results for $E^2_{0,0}$ are given in Refs.~\cite{HwangStable2026,ShiozakiAtiyahHirzebruch2023}.

In contrast, $E^\infty_{0,0}$ gives the stable adiabatic classification of phases obtained from atomic configurations.
For the symmetries considered here, each cell has internal symmetry class A, AI, or AII.
All three classes have a trivial stable classification in one spatial dimension~\cite{KitaevPeriodic2009,RyuTopological2010}.
Thus $E^1_{p,-1}=0$ and $E^3_{0,0}=E^2_{0,0}$.
In three spatial dimensions, the only remaining differential into $E^2_{0,0}$ is
\begin{align}
 d^3_{3,-2}:E^3_{3,-2}\longrightarrow E^3_{0,0}.
 \label{clsnew:d3}
\end{align}
Hence~\cite{ShiozakiGeneralized2023}
\begin{align}
 E^\infty_{0,0}\cong E^2_{0,0}/\im d^3_{3,-2}.
 \label{clsnew:stable-quotient-main}
\end{align}
In particular,
\begin{align}
 \mathcal A\sim_{\mathrm{st\,ad}}\mathcal B
 \quad\Longleftrightarrow\quad
 [\mathcal A]-[\mathcal B]\in\im d^3_{3,-2}.
 \label{clsnew:stable-test}
\end{align}
Equations~\eqref{clsnew:E2-stW} and \eqref{clsnew:stable-test} show that stable Wyckoff and stable adiabatic equivalence coincide for all atomic configurations if and only if $d^3_{3,-2}=0$.
In two or fewer dimensions, $d^3_{3,-2}$ is also absent, giving the following result.
\begin{thm}
In two or fewer spatial dimensions, stable adiabatic equivalence and stable Wyckoff equivalence of atomic configurations coincide.
\end{thm}

The third differential describes the difference in atomic configurations left by the creation, transport, and annihilation of two-dimensional topological phases.
The domain $E^3_{3,-2}$ of the third differential consists of candidate pumps that satisfy the consistency conditions on lower-dimensional cells.
A nonzero element of $\im d^3_{3,-2}$ is a configuration difference that cannot be removed by Wyckoff deformations but can be removed by a stable adiabatic deformation involving a pump.
The Chern-insulator pump in Sec.~\ref{sec:adia_1} is an explicit example of a pump of a two-dimensional topological phase
~\cite{ShiozakiGeneralized2023,WadaGeneral2024}.

The differential $d^3_{3,-2}$ measures the obstruction to realizing a pump throughout real space.
Thus $d^3_{3,-2}=0$ if there exists a model $V_\bk$ that represents the generator,
is invariant under continuous translations, and satisfies the magnetic space-group symmetry condition
\begin{align}
    u(g)V_\bk^{\phi_g}u(g)^\dagger=V_{\phi_gp_g\bk},\qquad g\in G.
\end{align}

\subsection{Symmetries that allow pumps}
\label{clsnew:source}

For magnetic space groups without pure time reversal (Types~I, III, and IV), two-dimensional phases at a generic point are labeled by the Chern number.
Both orientation-reversing spatial operations and antiunitary operations reverse the sign of the pump.
Therefore,
\begin{align}
 E^2_{3,-2}\cong
 \begin{cases}
  \Z,&\phi_g\det p_g=1\quad(\forall g\in G),\\
  0,&\exists g\in G:\ \phi_g\det p_g=-1
 \end{cases}.
 \label{clsnew:integer-source}
\end{align}
The generator of $\Z$ corresponds to a pump with Chern number one.
We say that $G$ preserves the spacetime orientation if $\phi_g\det p_g=1$ for every $g$.
Type~IV groups, for example, contain antiunitary translations, which violate $\phi_g\det p_g=1$.
Since $E^2_{1,-1}=0$ and space is three-dimensional, the domain also satisfies $E^3_{3,-2}=E^2_{3,-2}$.

For Type~II groups, which contain pure time reversal, the two-dimensional spinless classification is trivial.
For spinful electrons, the candidate is a $\Z_2$ pump associated with a two-dimensional class AII phase.

\subsection{Classification for spinful electrons}
\label{clsnew:spinful}

\begin{thm}
\label{thm:spinful_uniqueness}
For three-dimensional magnetic space groups with the electronic spin-$1/2$ factor system,
$d^3_{3,-2}=0$.
Consequently, stable adiabatic equivalence and stable Wyckoff equivalence of atomic configurations coincide.
\end{thm}

\noindent\textit{Proof.} 
The result is immediate when the domain is trivial, so we consider the cases with a nontrivial generator.
For groups of Types~I and III that preserve the spacetime orientation, every operation is a combination of rotations, $PT$, and translations.
On the spin-$1/2$ internal space, let rotations and $PT$ act as
\begin{align}
 u(R_{\hat{\bm n},\theta})
 &=e^{-i\theta\hat{\bm n}\cdot\bm\sigma/2},
 &u(PT)&=i\sigma_y.
\end{align}
Here $\bm\sigma$ denotes the vector of spin Pauli matrices, and translations act trivially on the internal degrees of freedom.
The unitary operator on $\R^3$,
\begin{align}
 V_{\bk}
 &=\frac{(|\bk|^2-m)I_2-i\bk\cdot\bm\sigma}
 {\sqrt{(|\bk|^2-m)^2+|\bk|^2}},\qquad
 \bk=-i\bm\nabla,\quad m>0,
 \label{clsnew:q}
\end{align}
respects rotations, $PT$, and translations and has three-dimensional winding number $W_3[V]=1$.
The unit Chern pump can therefore be realized globally, giving $d^3_{3,-2}=0$.

For Type~II groups, consider rotations, inversion, time reversal, and translations.
We use an internal space with spin $1/2$ and two orbitals.
Rotations act as above, while inversion and time reversal act as
\begin{align}
 u(I)&=\tau_z,
 &u(T)&=i\sigma_y.
\end{align}
Here $\bm\tau$ denotes the vector of Pauli matrices in orbital space.
The operator
\begin{align}
 V_{\bk}
 &=\frac{(|\bk|^2-m)\tau_z+\tau_x\bk\cdot\bm\sigma}
 {\sqrt{(|\bk|^2-m)^2+|\bk|^2}},\qquad
 \bk=-i\bm\nabla,\quad m>0,
 \label{clsnew:CII-unitary}
\end{align}
is also a symmetry-compatible unitary operator and represents the generator of the two-dimensional class AII insulator pump.
Thus $d^3_{3,-2}=0$.
\hfill$\square$\par

\begin{cor}
\label{cor:spinful_atomic_uniqueness}
For spinful magnetic space groups, an atomic configuration realizing a given band representation is unique up to stable Wyckoff equivalence.
\end{cor}

\subsection{Classification for spinless electrons}
\label{clsnew:spinless}

\subsubsection{Constraints on the monopole charge}
\label{clsnew:monopole-constraints}

Consider spinless systems with $E^3_{3,-2}\cong\Z$.
Suppose that the site-symmetry group at some point contains $222$, $\bar1'$, or $\bar4'$.
Under $222$, $\bar1'$, or $\bar4'$, points away from the center occur in sets of even cardinality.
Thus, bringing orbitals to the center or splitting orbitals away from the center changes the number of occupied orbitals at the center only by an even integer.
The parity of the charge at the center is preserved under stable Wyckoff equivalence.

To analyze a unit Chern insulator pump near a symmetry center, consider a Chern insulator layer with Chern number one on a sphere surrounding the center.
In a two-component Dirac description, the spinless symmetry algebra at a $222$, $\bar1'$, or $\bar4'$ center requires the monopole charge inside the sphere to be odd.
The magnetic flux through the sphere is therefore an odd multiple of the flux quantum.
Since the layer has Chern number one, the Hall response induces one unit of $U(1)$ charge per flux quantum, giving an odd induced charge on the spherical layer.

We derive the constraint on the monopole charge $m\in\Z$ using a Dirac model on a sphere enclosing a monopole.
In a fixed spinor basis, denote the wave functions in the northern and southern patches by
$\psi_N(\bx)$ and $\psi_S(\bx)$, respectively, with $\bx\in S^2$.
On the overlap of the two patches, we impose the gluing relation
\begin{align}
 \psi_N(\bx)=e^{im\phi}\psi_S(\bx),
 \label{clsnew:monopole-gluing}
\end{align}
where $\phi$ is the azimuthal angle of $\bx$.
The matrices $\bm\sigma$ below act on the two components of the Dirac model and do not represent the physical spin.
We impose the spinless symmetry algebra on the full transformations, including the action on coordinates and magnetic flux.
We denote the transformed wave function by $\psi'$ and require it to satisfy Eq.~\eqref{clsnew:monopole-gluing} as well.

For $PT$, define the northern component of $\psi'=PT\psi$ by
$\psi'_N(\bx)=i\sigma_y\psi_S(-\bx)^*$.
Since $-\bx$ has azimuthal angle $\phi+\pi$, the gluing relation fixes the southern component as
\begin{align}
 \psi'_S(\bx)
 &=e^{-im\phi}i\sigma_y\psi_S(-\bx)^*\notag\\
 &=(-1)^m i\sigma_y\psi_N(-\bx)^*.
 \label{clsnew:monopole-PT-sewing}
\end{align}
Applying $PT$ once more gives
\begin{align}
 \bigl((PT)^2\psi\bigr)_N(\bx)
 &=i\sigma_y\bigl[(-1)^m i\sigma_y\psi_N(\bx)^*\bigr]^*\notag\\
 &=(-1)^{m+1}\psi_N(\bx).
 \label{clsnew:monopole-PT-square}
\end{align}
The spinless condition $(PT)^2=1$ therefore requires $m$ to be odd.
The symmetries $\bar4'$ and $222$ also require odd $m$, as proved in Appendix~\ref{app:monopole-symmetries}.

For groups with a $222$, $\bar1'$, or $\bar4'$ center, a unit Chern pump therefore leaves an odd charge at the center.
Since a stable Wyckoff deformation cannot remove this odd charge,
\begin{align}
 d^3_{3,-2}(1)\ne0.
 \label{clsnew:local-nonzero}
\end{align}

Whether the fixed-point condition is also necessary is a separate question.
Indeed, in time-reversal-symmetric topological superconductors, symmetry can enforce vortex zero modes even when the vortex is not pinned to a particular high-symmetry point~\cite{OnoClassification2024}.
Thus a defect pinned to a high-symmetry point gives a sufficient condition for a nonzero higher differential $d^r$ ($r\geq2$), but not a necessary condition in general.
We next verify that the condition is also necessary for the symmetries considered here.

\subsubsection{Classification over all magnetic space groups}
\label{clsnew:enumeration}

First, we show that twice the unit pump can always be realized globally in spinless systems.
For a group that preserves the spacetime orientation, denote the matrix part of the spin-$1/2$ action in Eq.~\eqref{clsnew:q} by $u_g$ and the associated factor system by $z_{g,h}$.
Since $u_gu_h^{\phi_g}=z_{g,h}u_{gh}$, taking the tensor product with the complex-conjugate representation cancels the multiplier:
\begin{align}
 \widetilde u_g:=u_g\otimes u_g^*,\qquad
 \widetilde u_g\widetilde u_h^{\phi_g}=\widetilde u_{gh}.
 \label{clsnew:linear-double}
\end{align}
The matrices $\widetilde u_g$ therefore define a linear spinless action.
The two-copy Chern-insulator pump $V_{\bk}\otimes I_2$ respects the symmetry action defined by $\widetilde u_g$.
Since $W_3[V\otimes I_2]=2$, we obtain $2d^3_{3,-2}(1)=0$.

A nonzero third differential therefore requires $E^3_{3,-2}\cong\Z$ and $E^2_{0,0}$ to contain an element of order two.
The real-space $E^2$ data for all 1651 magnetic space groups~\cite{ShiozakiAtiyahHirzebruch2023} show that
230 groups have $E^3_{3,-2}\cong\Z$, and 137 of the 230 groups also have an element of order two in $E^2_{0,0}$.
Of these 137 groups, 134 have a point whose site-symmetry group contains $222$, $\bar1'$, or $\bar4'$ as a subgroup.
There are 230 such groups because, for each space group, the condition $\phi_g=\det p_g$ uniquely determines which operations are unitary and which are antiunitary.

The three remaining exceptions are
\begin{align}
 Pm',\qquad Pm'c'2_1,\qquad P\bar6'.
 \label{clsnew:three}
\end{align}
To establish the necessity of the fixed-point condition, it therefore suffices to show that $d^3_{3,-2}=0$ for $Pm'$, $Pm'c'2_1$, and $P\bar6'$.

\subsubsection{Models with unit winding number for the three exceptional groups}
\label{clsnew:three-details}
We construct an equivariant unitary with $W_3=1$ for each group in Eq.~\eqref{clsnew:three}.

We use the two-component model in Eq.~\eqref{clsnew:q} and construct an internal action with a trivial factor system for each group.
Since the model is invariant under continuous translations, it suffices to check the point-group actions.

For the magnetic point group $m'$ of $Pm'$, choose $u(m_y')=I_2$ for the generator $m_y'$.
Then
\begin{align}
 u(m_y')u(m_y')^*=I_2,\qquad
 u(m_y')V_{\bk}^*u(m_y')^\dagger=V_{-m_y\bk},
 \label{clsnew:Pm-model}
\end{align}
so both the group relation and the symmetry condition are satisfied.
The magnetic point group $m'm'2$ of $Pm'c'2_1$ has an additional generator $m_x'$.
Choosing $u(m_x')=\sigma_z$ gives
\begin{align}
 &u(m_x')u(m_x')^*=I_2,\qquad
 u(m_x')u(m_y')^*=u(m_y')u(m_x')^*,\notag\\
 &u(m_x')V_{\bk}^*u(m_x')^\dagger=V_{-m_x\bk}.
\end{align}
The group relations and symmetry conditions are again satisfied.

For $P\bar6'$, choose $u(\bar6')=i\sigma_x e^{-i\pi\sigma_z/3}$ for the magnetic point-group generator $\bar6'$.
Then
\begin{align}
 (u(\bar6')u(\bar6')^*)^3=I_2,\qquad
 u(\bar6')V_{\bk}^*u(\bar6')^\dagger=V_{-\bar6\bk}.
\end{align}
Thus both the group relation and the covariance condition are satisfied.
All three groups therefore admit a model with $W_3[V]=1$, and $d^3_{3,-2}=0$.

\subsubsection{Classification theorem}
The results above give the following theorem.

\begin{thm}
\label{thm:spinless_classification}
For a three-dimensional magnetic space group with a trivial factor system, $d^3_{3,-2}\ne0$ if and only if both of the following conditions hold:
\begin{enumerate}
 \item $\phi_g\det p_g=1$ for every $g\in G$.
 \item The site-symmetry group at some point contains $222$, $\bar1'$, or $\bar4'$ as a subgroup.
\end{enumerate}
These conditions hold for 134 groups: 22 of Type~I and 112 of Type~III. For these groups,
\begin{align}
 \im d^3_{3,-2}&\cong\Z_2,\notag\\
 \ker d^3_{3,-2}&=2\Z\subset E^3_{3,-2}\cong\Z.
 \label{clsnew:spinless-result}
\end{align}
For each of these 134 groups, there exist atomic configurations that are stably adiabatically equivalent but not stably Wyckoff equivalent.
For all other magnetic space groups, the two stable equivalence relations coincide.
\end{thm}

The 134 groups are listed in Appendix~\ref{clsnew:group-list}.

\begin{cor}
\label{cor:spinless_atomic_nonuniqueness}
For each of the 134 spinless magnetic space groups in Theorem~\ref{thm:spinless_classification},
there exist two atomic configurations that realize the same band representation but are not stably Wyckoff equivalent.
Thus a band representation does not, in general, uniquely determine the stable Wyckoff equivalence class of an atomic configuration that realizes it.
\end{cor}

\begin{cor}
\label{cor:spinless_path_constraints}
For each of the 134 groups in Theorem~\ref{thm:spinless_classification}, there exists a pair of atomic configurations $\mathcal A,\mathcal B$
that are stably adiabatically equivalent but not stably Wyckoff equivalent.
Any symmetric stable adiabatic path connecting $\mathcal A$ and $\mathcal B$ is a Chern pump that necessarily breaks time-reversal symmetry.
Moreover, if a local boundary termination of a semi-infinite system is fixed and the boundaries at both endpoints are also in the atomic limit,
the surface gap must close along the path.
\end{cor}

\noindent\textit{Proof.} 
Any unitary $V_\bk$ that defines an adiabatic path has odd winding number.
By the same argument as in Sec.~\ref{sec:sCS}, the second Chern number associated with the adiabatic path is therefore odd when both endpoints are atomic limits.
Consequently, the gap must close at the real-space surface.
\hfill$\square$\par

\section{Conclusion}
\label{sec:conclusion}

Even in a Wannierizable insulator, the atomic configuration defined by a Wannier basis need not be unique.
We distinguished Wyckoff equivalence, defined by continuous deformations of atomic configurations, from adiabatic equivalence, defined by continuous deformations of electronic states.
For each, we also introduced stable equivalence by allowing common occupied ancillas.

In the spinless $Fm'm'm'$ example, two stably Wyckoff-equivalent configurations are adiabatically inequivalent when the number of occupied bands is fixed at two.
The Euler class of a real rank-two bundle that descends to the symmetry quotient detects the obstruction to adiabatic equivalence.
This obstruction disappears upon adding occupied ancillas.

The spinless $P\overline1'$ example shows the converse distinction: two configurations are adiabatically equivalent but not stably Wyckoff equivalent.
The same occupied subspace admits Wannier bases belonging to different stable Wyckoff classes.
Thus topological invariants of the isomorphism class of a band representation do not uniquely determine the stable Wyckoff class of an atomic configuration.

The difference between the two stable equivalence relations is described by the image of the third differential
$d^3_{3,-2}:E^3_{3,-2}\to E^3_{0,0}$ in the real-space AHSS.
The differential $d^3_{3,-2}$ describes how a pump of a two-dimensional topological phase stably and adiabatically connects configurations that cannot be connected by a stable Wyckoff deformation.
Hence stable Wyckoff equivalence and stable adiabatic equivalence coincide when $d^3_{3,-2}=0$.

We determined $d^3_{3,-2}$ for spinful and spinless systems with three-dimensional magnetic space groups.
For spinful systems, $d^3_{3,-2}$ always vanishes, so a band representation uniquely determines the stable Wyckoff class of an atomic configuration.
For spinless systems, $d^3_{3,-2}$ is nonzero precisely for the 134 groups in Theorem~\ref{thm:spinless_classification}.
Each of the 134 groups admits a pair of configurations that are stably adiabatically equivalent but not stably Wyckoff equivalent.
Adding a common occupied ancilla then gives two stably Wyckoff-inequivalent configurations that realize the same band representation.

Any symmetric stable adiabatic path between these atomic limits, which are not stably Wyckoff equivalent, involves a Chern insulator pump.
If a local termination of a semi-infinite system is fixed and both endpoint boundaries are also in the atomic limit,
the surface gap must close along the path.

Extensions to spin space groups, in which spatial and spin rotations act independently, and to general factor systems remain for future work.

\begin{acknowledgments}
\label{sec:acknowledgments}

The author thanks Shuichi~Murakami, Seishiro~Ono, and Hidetoshi~Wada for helpful discussions.
Codex (OpenAI GPT-6 Astra) was used to assist with theoretical discussions,
the organization and checking of derivations, literature searches, manuscript drafting and editing,
and the preparation of computational code and figures.
The author is responsible for the content and conclusions of the paper.
This work was supported by JSPS KAKENHI Grant Nos.~JP22H05118, JP26H01305, and JP26K00629.

\end{acknowledgments}

\appendix

\section{Gapped Dirac models for $pmm$ and $P222$}
\label{app:examples_intertwiners}
We give models that connect the two configurations in Secs.~\ref{ex:pmm_stW} and \ref{ex:P222_stW} without adding occupied ancillas.
Using a nonperiodic Bloch basis that includes the orbital-position phases, we construct
$V_\bk:\mathcal H_{\mathcal B,\bk}\to\mathcal H_{\mathcal A,\bk}$.
Let $D_j^{\mathcal A}$ and $D_j^{\mathcal B}$ denote the basis transition matrices under reciprocal-lattice translations.
The gluing and symmetry conditions are
\begin{align}
    V_{\bk+2\pi\mathbf e_j}
    &=D_j^{\mathcal A}V_\bk(D_j^{\mathcal B})^\dagger,\notag\\
    u^{\mathcal A}(g)V_\bk&=V_{g\bk}u^{\mathcal B}(g),
\end{align}
where $\mathbf e_j$ is the unit vector in the $j$th direction and $g$ is a point-group generator.
For an invertible $V_\bk$, polar decomposition gives the unitary matrix
$V_\bk(V_\bk^\dagger V_\bk)^{-1/2}$, which satisfies the gluing and symmetry conditions.
Below, $X,Y,Z$ denote Pauli matrices, and we abbreviate
$c_j:=\cos(k_j/2)$ and $s_j:=\sin(k_j/2)$.

\subsection{$pmm$: a model with four occupied bands}
With the sites ordered as $(a,b,c,d)$, the mirror matrices of the two configurations are
\begin{align}
    u^{\mathcal A}(m_x)&=-u^{\mathcal B}(m_x)=-I_2\otimes Z,\notag\\
    u^{\mathcal A}(m_y)&=-u^{\mathcal B}(m_y)=I_2\otimes Z.
\end{align}
The orbital positions of $\mathcal A$ and $\mathcal B$ coincide, so the basis transition matrices also coincide:
\begin{align}
    D_x^{\mathcal A}=D_x^{\mathcal B}&=Z\otimes Z,\notag\\
    D_y^{\mathcal A}=D_y^{\mathcal B}&=Z\otimes I_2.
\end{align}
Set
\begin{align}
    V_\bk
    =c_x(I_2\otimes X)+c_y(X\otimes Y)+s_xs_y(X\otimes Z).
\end{align}
Then $V_\bk^\dagger V_\bk
    =(c_x^2+c_y^2+s_x^2s_y^2)I_4
    =(1+c_x^2c_y^2)I_4\ge I_4$,
so $V_\bk$ is invertible at every momentum.

\subsection{$P222$: a model with eight occupied bands}
Configuration $\mathcal A$ has two $s$ orbitals at each even vertex, and $\mathcal B$ has two at each odd vertex.
Order the sites as $(000,\allowbreak101,\allowbreak110,\allowbreak011)$ for $\mathcal A$
and $(111,\allowbreak010,\allowbreak001,\allowbreak100)$ for $\mathcal B$.
Of the three tensor factors, the first two distinguish the four sites, and the last distinguishes the two orbitals at each site.
In the basis ordered by site and orbital, the twofold-rotation matrices are
$u^{\mathcal A}(C_{2j})=u^{\mathcal B}(C_{2j})=I_8$, and the basis transition matrices are
\begin{align}
    D_x^{\mathcal A}=-D_x^{\mathcal B}&=Z\otimes Z\otimes I_2,\notag\\
    D_y^{\mathcal A}=-D_y^{\mathcal B}&=Z\otimes I_2\otimes I_2,\notag\\
    D_z^{\mathcal A}=-D_z^{\mathcal B}&=I_2\otimes Z\otimes I_2.
\end{align}
Set
\begin{align}
    V_\bk
    ={}&c_x(X\otimes X\otimes X)
        +c_y(I_2\otimes Y\otimes X)\notag\\
       &+c_z(X\otimes Z\otimes X)
        +s_xs_ys_z(I_2\otimes I_2\otimes Z).
\end{align}
Then $V_\bk^\dagger V_\bk=(c_x^2+c_y^2+c_z^2+s_x^2s_y^2s_z^2)I_8=(1+c_x^2c_y^2s_z^2+c_x^2c_z^2+c_y^2c_z^2)I_8\ge I_8$,
so $V_\bk$ is invertible at every momentum.

\section{Odd monopole charge enforced by spinless \texorpdfstring{$\bar4'$ and $222$}{-4' and 222} symmetries}
\label{app:monopole-symmetries}

We use the spherical Dirac model in Sec.~\ref{clsnew:monopole-constraints}, with monopole charge $m\in\Z$.
We impose the gluing relation in Eq.~\eqref{clsnew:monopole-gluing} on both the original wave functions $\psi_N,\psi_S$
and the transformed wave functions $\psi'_N,\psi'_S$.
Here $\phi$ is the azimuthal angle, and $\bm\sigma$ are the Pauli matrices of the Dirac spinor in a fixed basis.
The spinless symmetry algebra is imposed on the full transformations, including the action on coordinates and magnetic flux.

\subsection{The case of \texorpdfstring{$\bar4'$}{-4'}}
\label{app:monopole-S4}

For $\bar4'$, take the spatial action $s_4(x,y,z)=(y,-x,-z)$ and set
$u(s_4)=e^{-i\pi\sigma_z/4}i\sigma_y$.
Defining the northern component of $\psi'=\bar4'\psi$ also fixes the southern component:
\begin{align}
 \psi'_N(\bx)&=u(s_4)\psi_S(s_4^{-1}\bx)^*,\notag\\
 \psi'_S(\bx)&=e^{im\pi/2}u(s_4)\psi_N(s_4^{-1}\bx)^*.
 \label{clsnew:monopole-S4-sewing}
\end{align}
Here we used the fact that $s_4^{-1}\bx$ has azimuthal angle $\phi+\pi/2$.
Since $u(s_4)u(s_4)^*=i\sigma_z$, applying the operation twice and four times gives
\begin{align}
 \bigl((\bar4')^2\psi\bigr)_N(\bx)
 &=e^{-im\pi/2}i\sigma_z\psi_N(c_2^z\bx),\notag\\
 \bigl((\bar4')^4\psi\bigr)_N(\bx)
 &=(-1)^{m+1}\psi_N(\bx).
 \label{clsnew:monopole-S4-fourth}
\end{align}
Thus the spinless condition $(\bar4')^4=1$ also requires $m$ to be odd.

\subsection{The case of \texorpdfstring{$222$}{222}}
\label{app:monopole-222}

For $222$, take $c_2^x(x,y,z)=(x,-y,-z)$ and $c_2^y(x,y,z)=(-x,y,-z)$,
and use $u(c_2^x)=-i\sigma_x$ and $u(c_2^y)=-i\sigma_y$.
Choose the overall phase so that, for $\psi'=c_2^x\psi$,
\begin{align}
 \psi'_N(\bx)&=e^{im\pi/2}u(c_2^x)\psi_S(c_2^x\bx),\notag\\
 \psi'_S(\bx)&=e^{im\pi/2}u(c_2^x)\psi_N(c_2^x\bx).
 \label{clsnew:monopole-C2x-sewing}
\end{align}
For $\psi'=c_2^y\psi$, we instead have
\begin{align}
 \psi'_N(\bx)&=u(c_2^y)\psi_S(c_2^y\bx),\notag\\
 \psi'_S(\bx)&=(-1)^m u(c_2^y)\psi_N(c_2^y\bx).
 \label{clsnew:monopole-C2y-sewing}
\end{align}
In both cases, the expression in the southern patch follows from the gluing relation.
Applying the two operations in opposite orders gives
\begin{align}
 (c_2^xc_2^y\psi)_N(\bx)
 &=e^{im\pi/2}(-1)^m u(c_2^x)u(c_2^y)\psi_N(c_2^z\bx),\notag\\
 (c_2^yc_2^x\psi)_N(\bx)
 &=e^{im\pi/2}u(c_2^y)u(c_2^x)\psi_N(c_2^z\bx).
 \label{clsnew:monopole-222-orders}
\end{align}
Since $u(c_2^x)u(c_2^y)=-u(c_2^y)u(c_2^x)$, the full operations commute only if $(-1)^{m+1}=1$, or equivalently if $m$ is odd.
For odd $m$, each twofold rotation also squares to the identity, so the spinless $222$ algebra is satisfied.

\section{The 134 magnetic space groups}
\label{clsnew:group-list}

\begingroup\raggedright
\noindent\textbf{Type I (22 groups).}
$\mathrm{P222}$,
$\mathrm{C222}$,
$\mathrm{F222}$,
$\mathrm{I222}$,
$\mathrm{P422}$,
$\mathrm{P42_12}$,
$\mathrm{P4_222}$,
$\mathrm{P4_22_12}$,
$\mathrm{I422}$,
$\mathrm{I4_122}$,
$\mathrm{P622}$,
$\mathrm{P6_222}$,
$\mathrm{P6_422}$,
$\mathrm{P23}$,
$\mathrm{F23}$,
$\mathrm{I23}$,
$\mathrm{P432}$,
$\mathrm{P4_232}$,
$\mathrm{F432}$,
$\mathrm{F4_132}$,
$\mathrm{I432}$,
$\mathrm{I4_132}$.

\noindent\textbf{Type III (112 groups).}\par\noindent
$\mathrm{P\bar{1}'}$,
$\mathrm{P2/m'}$,
$\mathrm{P2_1/m'}$,
$\mathrm{C2/m'}$,
$\mathrm{P2/c'}$,
$\mathrm{P2_1/c'}$,
$\mathrm{C2/c'}$,
$\mathrm{Pm'm'm'}$,
$\mathrm{Pn'n'n'}$,
$\mathrm{Pc'c'm'}$,
$\mathrm{Pb'a'n'}$,
$\mathrm{Pm'm'a'}$,
$\mathrm{Pn'n'a'}$,
$\mathrm{Pm'n'a'}$,
$\mathrm{Pc'c'a'}$,
$\mathrm{Pb'a'm'}$,
$\mathrm{Pc'c'n'}$,
$\mathrm{Pb'c'm'}$,
$\mathrm{Pn'n'm'}$,
$\mathrm{Pm'm'n'}$,
$\mathrm{Pb'c'n'}$,
$\mathrm{Pb'c'a'}$,
$\mathrm{Pn'm'a'}$,
$\mathrm{Cm'c'm'}$,
$\mathrm{Cm'c'a'}$,
$\mathrm{Cm'm'm'}$,
$\mathrm{Cc'c'm'}$,
$\mathrm{Cm'm'a'}$,
$\mathrm{Cc'c'a'}$,
$\mathrm{Fm'm'm'}$,
$\mathrm{Fd'd'd'}$,
$\mathrm{Im'm'm'}$,
$\mathrm{Ib'a'm'}$,
$\mathrm{Ib'c'a'}$,
$\mathrm{Im'm'a'}$,
$\mathrm{P\bar{4}'}$,
$\mathrm{I\bar{4}'}$,
$\mathrm{P4/m'}$,
$\mathrm{P4_2/m'}$,
$\mathrm{P4/n'}$,
$\mathrm{P4_2/n'}$,
$\mathrm{I4/m'}$,
$\mathrm{I4_1/a'}$,
$\mathrm{P\bar{4}'2m'}$,
$\mathrm{P\bar{4}'2c'}$,
$\mathrm{P\bar{4}'2_1m'}$,
$\mathrm{P\bar{4}'2_1c'}$,
$\mathrm{P\bar{4}'m'2}$,
$\mathrm{P\bar{4}'c'2}$,
$\mathrm{P\bar{4}'b'2}$,
$\mathrm{P\bar{4}'n'2}$,
$\mathrm{I\bar{4}'m'2}$,
$\mathrm{I\bar{4}'c'2}$,
$\mathrm{I\bar{4}'2m'}$,
$\mathrm{I\bar{4}'2d'}$,
$\mathrm{P4/m'm'm'}$,
$\mathrm{P4/m'c'c'}$,
$\mathrm{P4/n'b'm'}$,
$\mathrm{P4/n'n'c'}$,
$\mathrm{P4/m'b'm'}$,
$\mathrm{P4/m'n'c'}$,
$\mathrm{P4/n'm'm'}$,
$\mathrm{P4/n'c'c'}$,
$\mathrm{P4_2/m'm'c'}$,
$\mathrm{P4_2/m'c'm'}$,
$\mathrm{P4_2/n'b'c'}$,
$\mathrm{P4_2/n'n'm'}$,
$\mathrm{P4_2/m'b'c'}$,
$\mathrm{P4_2/m'n'm'}$,
$\mathrm{P4_2/n'm'c'}$,
$\mathrm{P4_2/n'c'm'}$,
$\mathrm{I4/m'm'm'}$,
$\mathrm{I4/m'c'm'}$,
$\mathrm{I4_1/a'm'd'}$,
$\mathrm{I4_1/a'c'd'}$,
$\mathrm{P\bar{3}'}$,
$\mathrm{R\bar{3}'}$,
$\mathrm{P\bar{3}'1m'}$,
$\mathrm{P\bar{3}'1c'}$,
$\mathrm{P\bar{3}'m'1}$,
$\mathrm{P\bar{3}'c'1}$,
$\mathrm{R\bar{3}'m'}$,
$\mathrm{R\bar{3}'c'}$,
$\mathrm{P6/m'}$,
$\mathrm{P6_3/m'}$,
$\mathrm{P6/m'm'm'}$,
$\mathrm{P6/m'c'c'}$,
$\mathrm{P6_3/m'c'm'}$,
$\mathrm{P6_3/m'm'c'}$,
$\mathrm{Pm'\bar{3}'}$,
$\mathrm{Pn'\bar{3}'}$,
$\mathrm{Fm'\bar{3}'}$,
$\mathrm{Fd'\bar{3}'}$,
$\mathrm{Im'\bar{3}'}$,
$\mathrm{Pa'\bar{3}'}$,
$\mathrm{Ia'\bar{3}'}$,
$\mathrm{P\bar{4}'3m'}$,
$\mathrm{F\bar{4}'3m'}$,
$\mathrm{I\bar{4}'3m'}$,
$\mathrm{P\bar{4}'3n'}$,
$\mathrm{F\bar{4}'3c'}$,
$\mathrm{I\bar{4}'3d'}$,
$\mathrm{Pm'\bar{3}'m'}$,
$\mathrm{Pn'\bar{3}'n'}$,
$\mathrm{Pm'\bar{3}'n'}$,
$\mathrm{Pn'\bar{3}'m'}$,
$\mathrm{Fm'\bar{3}'m'}$,
$\mathrm{Fm'\bar{3}'c'}$,
$\mathrm{Fd'\bar{3}'m'}$,
$\mathrm{Fd'\bar{3}'c'}$,
$\mathrm{Im'\bar{3}'m'}$,
$\mathrm{Ia'\bar{3}'d'}$.
\par\endgroup

% REVTeX selects apsrev4-2 and includes article titles.
\bibliography{paper_refs}

\end{document}